\documentclass[apj,twocolumn]{openjournal}
\usepackage{amsmath}
\usepackage{booktabs}

\usepackage[dvipsnames]{xcolor} 
\usepackage[breaklinks,colorlinks,citecolor=blue,urlcolor=blue]{hyperref}

\usepackage{orcidlink}
\usepackage{subfigure,comment}
\usepackage{tablefootnote}

\renewcommand{\arraystretch}{1.2}  
\usepackage{listings}
\usepackage{color}
\definecolor{dkgreen}{rgb}{0,0.6,0}
\definecolor{gray}{rgb}{0.5,0.5,0.5}
\definecolor{mauve}{rgb}{0.58,0,0.82}
\definecolor{golden}{rgb}{0.86,0.65,0.01}
\defcitealias{Xing2023}{X23}
\defcitealias{Jeena2024}{J24}

\usepackage{soul}
\usepackage{amsmath}
\usepackage{xspace}
\usepackage{graphicx}
\usepackage{enumitem}
\usepackage{amssymb}
\usepackage{xifthen}
\usepackage{hyperref}
\usepackage[normalem]{ulem}
\usepackage{multirow}

\newcommand{\code}[1]{\texttt{#1}\xspace}

\newcommand{\unit}[1]{\ensuremath{\mathrm{\,#1}}\xspace}
\newcommand{\feh}{\unit{[Fe/H]}}

\newcommand{\xfe}[1]{\unit{[#1/Fe]}}
\newcommand{\xh}[1]{\unit{[#1/H]}}
\newcommand{\vt}{\ensuremath{v_t}\xspace}
\newcommand{\teff}{\ensuremath{T_\mathrm{eff}}\xspace}
\newcommand{\logg}{\ensuremath{\log\,g}\xspace}
\newcommand{\alphafe}{\unit{[\alpha/Fe]}}

\begin{document}


\author{Pierre N. Thibodeaux\,\orcidlink{0000-0002-3867-3927}$^{1,2}$}
\author{Alexander P. Ji\,\orcidlink{0000-0002-4863-8842}$^{1,2,3}$}
\author{Nicholas Storm\,\orcidlink{0000-0002-5259-3974}$^{4,5}$}
\author{Maria Bergemann\,\orcidlink{0000-0002-9908-5571}$^4$}
\author{Rana Ezzeddine\,\orcidlink{0000-0002-8504-8470}$^6$}
\author{Yuan-sen Ting\,\orcidlink{0000-0001-5082-9536}$^{4,7}$}
\author{Andrew R. Casey\,\orcidlink{0000-0003-0174-0564}$^{8,9}$}
\author{Emily Griffith\,\orcidlink{0000-0001-9345-9977}$^{10}$}
\author{Jos\'e Fern\'andez-Trincado\,\orcidlink{0000-0003-3526-5052}$^{11}$}
\author{Guilherme Limberg\,\orcidlink{0000-0002-9269-8287}$^{1,2}$}
\author{Szabolcs M\'esz\'aros\,\orcidlink{0000-0001-8237-5209}$^{12,13}$}
\author{Amaya Sinha\,\orcidlink{0009-0005-0182-7186}$^{14,17}$}
\author{Danny Horta\,\orcidlink{0000-0003-1856-2151}$^{15}$}
\author{Andrew K. Saydjari\,\orcidlink{0000-0002-6561-9002}$^{16}$}
\author{Joel Brownstein\,\orcidlink{0000-0002-8725-1069}$^{14}$}

\affiliation{$^1$Department of Astronomy \& Astrophysics, University of Chicago, 5640 S Ellis Avenue, Chicago, IL 60637, USA}
\affiliation{$^2$Kavli Institute for Cosmological Physics, University of Chicago, Chicago, IL 60637, USA}
\affiliation{$^3$Joint Institute for Nuclear Astrophysics}
\affiliation{$^4$Max-Planck-Institut für Astronomie, Königstuhl 17, D-69117 Heidelberg, Germany}
\affiliation{$^5$Heidelberg University, Grabengasse 1, D-69117 Heidelberg, Germany}
\affiliation{$^6$University of Florida, Department of Astronomy, 211 Bryant Space Science Center}
\affiliation{$^7$Department of Astronomy, The Ohio State University, Columbus, OH 43210, USA}
\affiliation{$^8$Center for Computational Astrophysics, Flatiron Institute, 162 5th Ave, New York, NY 10010, USA}
\affiliation{$^9$School of Physics and Astronomy, Monash University, VIC 3800, Australia}
\affiliation{$^{10}$Center for Astrophysics and Space Astronomy, Department of Astrophysical and Planetary Sciences, University of Colorado, 389 UCB, Boulder, CO 80309-
0389, USA}
\affiliation{$^{11}$Centro de investigaci\'on en Astronom\'ia, Facultad de Ingenier\'ia, Ciencia y Tecnolog\'ia, Universidad Bernardo O’Higgins, Av. Viel 1497, Santiago, 8370993, Chile}
\affiliation{$^{12}$ELTE Eötvös Loránd University, Gothard Astrophysical Observatory, 9700 Szombathely, Szent Imre H. st. 112, Hungary}
\affiliation{$^{13}$HUN-REN CSFK, Konkoly Observatory, Konkoly Thege Mikl\'os \'ut 15-17, Budapest, 1121, Hungary}
\affiliation{$^{14}$Department of Physics and Astronomy, University of Utah, 270 S. 1400 E. \#E2108, Salt Lake City, UT 84112, USA}
\affiliation{$^{15}$Institute for Astronomy, University of Edinburgh, Royal Observatory, Blackford Hill, Edinburgh, EH9 3HJ, UK}
\affiliation{$^{16}$ Department of Astrophysical Sciences, Princeton University, Princeton, NJ 08544, USA}
\affiliation{$^{17}$ Space Telescope Science Institute, 3700 San Martin Drive, Baltimore, MD 21218, USA}

\email{Corresponding author: pthibodeaux@uchicago.edu}
\title{Payne4GAIN: NLTE Corrections for Red Giants in Milky Way Mapper using H-band Neural Network Emulators}

\begin{abstract}
The majority of spectroscopic surveys assume local thermodynamic equilibrium (LTE) during the modeling of stellar spectra. This assumption begins to break down for luminous stars, like the red giants targeted by SDSS-V's Milky Way Mapper Survey in its Galactic Genesis program. In this work, we present non-LTE (NLTE) abundances for 360,000 red giant stars in Milky Way Mapper DR19, from infrared APOGEE spectra. We generate NLTE spectra using precomputed departure coefficient grids for Na, Mg, Si, Al, Ca, Ti, Mn, and Ni. To fit APOGEE spectra at scale, we train neural network emulators (NNEs) to synthesize LTE and NLTE H-band spectra. After verifying that the NNEs are accurate, we fit the APOGEE spectra with ASPCAP results that fall within the same parameter range as the training data. We find strong NLTE effects on the order of 0.1\,dex for Al, Mn, and Ti, and smaller effects for Si and Ni. We provide a catalog of the results of our LTE and NLTE fits, as well as NLTE-corrected ASPCAP abundances using a polynomial fit correction.

\keywords{line: formation --- line: profiles ---radiative transfer --- stars: atmospheres --- stars: abundances --- techniques: spectroscopic --- Galaxy: abundances}
\end{abstract}

\maketitle
\section{Introduction}
\label{sec:intro}

The Milky Way Mapper (MWM) survey of the fifth generation of the Sloan Digital Sky Survey \citep[SDSS-V][]{kollmeier_sloan_2026} seeks to characterize the Milky Way with high-resolution spectra from across the entire sky. The primary program of MWM, Galactic Genesis, collects $R=$22,500 stellar spectra in the H-band between 15,000 and 17,000\AA\, using the APOGEE spectrographs \citep{wilson_apache_2019} on the SDSS telescope at Apache Point Observatory \citep[in New Mexico,][]{gunn_25_2006} and the Irénée du Pont telescope at Las Campanas Observatory \citep[in Chile,][]{bowen_optical_1973}. Galactic Genesis is a continuation of the Apache Point Observatory Evolution Experiment \citep[APOGEE;][]{majewski_apache_2017} from SDSS-III and SDSS-IV. Galactic Genesis targets luminous red giant stars to probe the Milky Way's bulge and disk \citep{zasowski_target_2017, sdss_collaboration_nineteenth_2025}. 

Red giants are among the brightest stars detectable, allowing us to obtain high resolution spectra at great distances. To understand the chemical and dynamical evolution of our Galaxy, it is important to acquire accurate, precise, and consistent stellar parameters and abundances of its stars \citep{freeman_new_2002,ivezic_galactic_2012,bland-hawthorn_galaxy_2016}. 

The accuracy of stellar labels is generally limited by the ability to model stellar spectra using physical simulations \citep{ness_homogeneity_2022}. There are many simplifying assumptions made when generating stellar spectra, such as 1D geometry, hydrostatic and radiative equilibrium, and more. However, one major assumption to the derivation of abundances is local thermodynamic equilibrium (LTE) for the species responsible for absorption features in stellar spectra. Under LTE, the local temperature uniquely determines the level populations of atoms responsible for line formation \citep{asplund_new_2005}.

For hot and/or luminous stars, radiative and collisional processes can violate LTE by altering atomic level populations, and require additional calculation to determine non-LTE (NLTE) level populations \citep{allende_prieto_solar_2016,jofre_accuracy_2019, bergemann_non-local_2017,bergemann_3d_2025}. This is a necessary consideration for O-stars \citep{lanz_grid_2003}, but for red giants, the effect of NLTE is less pronounced and often neglected. However, as the resolution and signal-to-noise ratio (SNR) of spectroscopic studies have increased, NLTE has become a necessary consideration for obtaining accurate stellar abundances \citep{amarsi_galah_2020, storm_observational_2025}. With 650,000 APOGEE spectra in DR17 \citep{abdurrouf_seventeenth_2022}, $\sim 1$ million in MWM DR19 \citep{meszaros_sdss-v_2025}, and a projection of over 3 million spectra by the end of SDSS-V \citep{sdss_collaboration_nineteenth_2025}, systematic abundance biases that may be caused by excluding NLTE become increasingly important.

A majority of existing NLTE literature focuses on fitting optical spectra. In this regime, it has been found that fitting in NLTE, compared to LTE, can contribute to differences in elemental abundances up to several tenths of a dex \citep[e.g.,][]{bergemann_non-local_2017}. There are also discrepancies between spectroscopic and photometrically derived values of stellar parameters \citep{gonzalez_hernandez_new_2009,ezzeddine_r-process_2020} that can be alleviated with NLTE for iron \citep{bergemann_non-lte_2012, lind_non-lte_2012,li_lotus_2023}. 

Given the profound effect of NLTE on optical spectra, it is natural to expect NLTE effects in APOGEE H-band (15,000--17,000\AA) spectra. In Milky Way Mapper, stellar labels are derived from APOGEE spectra using the APOGEE Stellar Parameter and Chemical Abundance Pipeline \citep[ASPCAP,][]{garcia_perez_aspcap_2016,jonsson_apogee_2020,meszaros_sdss-v_2025}. ASPCAP makes use of large grids of synthetic spectra to fit features in the spectra. In DR17, the spectral grids included Na, Mg, K, and Ca in NLTE. This is discussed in \citet{osorio_nlte_2020}, who found that NLTE corrections were $<0.10$ for all studied species. However, other analyses \citep[e.g.][]{nordlander_non-lte_2017, feltzing_metal-weak_2023,grilo_chemical_2024} have claimed larger corrections for other elements, like Al and Mn. 

To further the study of NLTE in the H-band, we employ NLTE calculations tested on red giants in other studies \citep[e.g.][]{ji_nearly_2025,storm_observational_2026} to fit  red giant spectra in MWM. In our analysis, we include 1D NLTE for 8 elements: Sodium, Magnesium, Aluminum, Silicon, Calcium, Titanium, Manganese, and Nickel. These elements were selected due to a combination of being important for interpreting MWM chemical evolution \citep[e.g.][]{limberg_reconstructing_2022,weinberg_chemical_2019, hasselquist_apogee_2021, griffith_similarity_2021} and having available NLTE information. As such, we focus on the effects that NLTE has on their chemical abundance trends by fitting spectra in LTE and NLTE and directly comparing the differences. As we explain in Section \ref{subsec:nltemethods}, we do not treat Iron in NLTE in this analysis.

This work is structured as follows: in Section \ref{sec:methods}, we discuss the process by which we generate LTE and NLTE H-band spectra, train neural network emulators (NNEs) to reproduce them, following a version of the \emph{Payne} \citep{ting_payne_2019}. We fit 360,000 spectra from MWM DR19, the most up-tp-date public sample, as no new APOGEE data are released in DR20. In Section \ref{sec:validation}, we validate the NNEs by demonstrating that our LTE NNE accurately reproduces LTE spectral synthesis and broadly agrees with results from ASPCAP. The only difference between our two Paynes is the presence/absence of NLTE, so we are able to isolate the abundance differences caused by the inclusion of NLTE. In Section \ref{sec:differences}, we present the results of our LTE and NLTE fits and derive NLTE-LTE corrections as functions of ASPCAP stellar parameters. We discuss the NLTE abundances in detail in Section \ref{sec:discussion}. We refer to our method as \code{Payne4GAIN} (Giants in APOGEE Including NLTE, also abbreviated P4G), and the results of our fits are included in the release of DR20 as a value added catalogue (VAC), which we discuss in Section \ref{sec:conclusion}.

\begin{figure*}
    \centering
    \includegraphics[width=\linewidth]{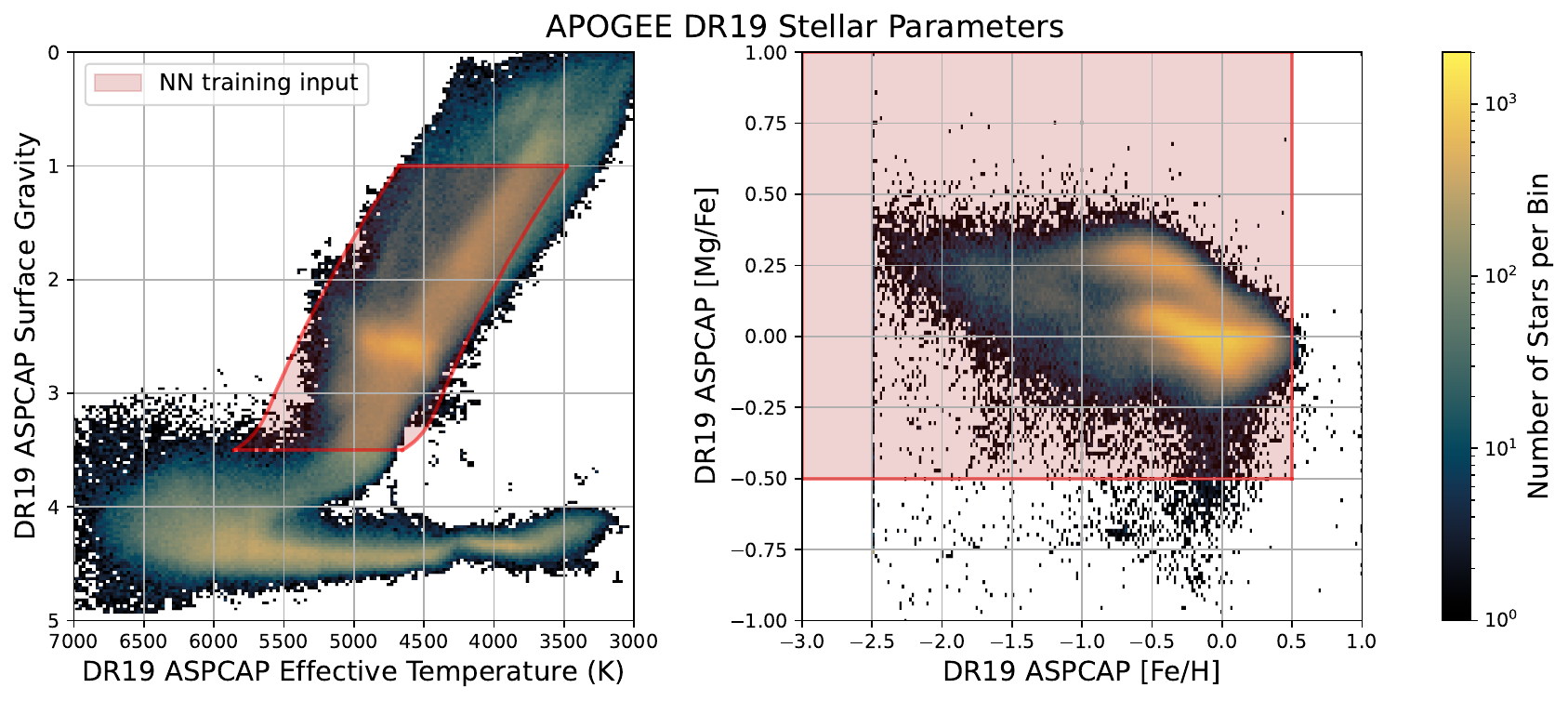}
    \caption{The Kiel diagram and the \alphafe vs \feh plot using the ASPCAP DR19 parameters. All values displayed here are raw, before calibration. The red shaded box indicates the range of the training inputs to our Paynes. We identify \alphafe with \xfe{Mg} for our NN for this plot.}
    \label{fig:dr19}
\end{figure*}

\section{Methods}
\label{sec:methods}

\subsection{Synthetic spectra in LTE and NLTE}
\label{subsec:nltemethods}

To generate NLTE spectra, we use the \code{TSFitPy} wrapper\footnote{\url{github.com/TSFitPy-developers/TSFitPy}} \citep{storm_observational_2023} of \code{Turbospectrum NLTE} \citep{gerber_non-lte_2023}, which is an updated version of \code{Turbospectrum} \citep{alvarez_near-infrared_1998,plez_turbospectrum_2012} that can use precomputed departure coefficients for multi-element synthesis\footnote{using a trace element approximation, where each element is calculated independently and assumed not to affect other elements.}. We use the standard MARCS model atmospheres \citep{gustafsson_grid_2008} and the atomic and molecular linelists developed for APOGEE \citep{shetrone_sdss-iii_2015,smith_apogee_2021}. The linelists are based on atomic data from VALD \citep{kupka_vald-2_1999,ryabchikova_major_2015} and NIST \citep{a_kramida_nist_2024} but had oscillator strengths adjusted to better match spectra of the Sun and Arcturus, following \citet{bizyaev_astrolines_2015}. Because this empirical tuning was performed in LTE, adding NLTE may result in some systematically biased line syntheses.

In LTE, the populations of molecules and ions are entirely specified with the Boltzmann \citep[orig. 1872]{boltzmann_weitere_1970} and Saha equations \citep{saha_physical_1921}. However, in NLTE, radiative and collisional rates (theoretical and experimental) must be considered. With a model atom that describes the energy states of a given element and a model stellar atmosphere, one can solve the statistical equilibrium equations, resulting in \emph{NLTE departure coefficients}, the ratio of each state's NLTE population to its LTE population.

The departure coefficients\footnote{\url{https://keeper.mpdl.mpg.de/d/6eaecbf95b88448f98a4/}} adopted in this analysis were generated using \code{MULTI} (\citealp{carlsson_computer_1986}, updated in \citealp{bergemann_observational_2019} and \citealp{gallagher_observational_2020}). These authors generated tables of departure coefficients for a range of MARCS model atmospheres \citep{gustafsson_grid_2008}. For the elements in our analysis, the model atoms used were: Na \citep{ezzeddine_empirical_2018}; Mg \citep{bergemann_non-local_2017}; Al \citep{ezzeddine_empirical_2018}; Si \citep{bergemann_red_2013,magg_observational_2022}; Ca \citep{mashonkina_influence_2017, semenova_gaia-eso_2020};  Ti \citep{bergemann_ionization_2011}; Mn \citep{bergemann_observational_2019}; and Ni \citep{bergemann_solar_2021,voronov_inelastic_2022}. 
The departure coefficient grids extend between \teff=2500—8000\,K, \logg=0-5.5, and \feh=$-5$ to +1, though there are gaps at the lowest metallicities ($<-2.0$) around \logg $\sim 0$, where the MARCS models fail to converge. This is the primary reason that our analysis ends at \logg=1, instead of extending to 0 like ASPCAP.

To make use of the departure coefficient grids, the upper and lower states of each atomic line must be matched to the appropriate states of the respective model atom. We used a routine from \code{TSFitPy} to match the APOGEE line list to the model atom states. This routine takes the excitation potential of each line as the lower state, calculates the energy of the upper state using the wavelength, and matches these states to the closest states of the model atoms. The matching procedure is necessarily approximate and sometimes matches lines to an incorrect adjacent state. However, we check several lines in detail\footnote{we chose the most sensitive 3 lines for each element, according to the windows from \citet{jonsson_apogee_2020}.}, manually matching them to the correct states, and find no significant differences in the resulting abundances.

We originally attempted to include NLTE Fe, but the Fe model atom \citep{bergemann_non-lte_2012, semenova_gaia-eso_2020} combines atomic levels for computational tractability. Because it misses states relevant to infrared lines, it produces unphysical spectra, so we exclude NLTE Fe in our analysis. Incorporating NLTE Fe remains a priority for future work, but doing so will require generating NLTE departure coefficients with a more complete Fe model atom.

\subsection{Neural Network Training}
The most basic version of spectral fitting involves using a synthesizer to generate a synthetic spectrum at some stellar labels, then iteratively adjusting the labels to optimize the $\chi^2$ with respect to some observed spectrum. Because we intend to fit hundreds of thousands of spectra, and spectrum generation takes at least a second even in LTE, it is not feasible to use this approach. ASPCAP accelerates fitting by interpolating pre-computed spectra, but this method is not compatible with a high-dimensional parameter space. We turn to machine learning to quickly fit stellar spectra with many labels.

Much work has been done to fit APOGEE spectra with neural networks. Discriminative models, which determine stellar labels from spectra \citep{fabbro_application_2018,leung_deep_2019} scale well, but can result in correlated parameters, attenuation-biased fits, and reproduce the biases of the pipeline labels they were trained on \citep{ting_why_2025,hogg_is_2024}. Generative models, which generate spectra at given stellar labels, have also been employed many times for APOGEE spectra, both data-driven \citep[e.g.][]{ness_cannon_2015,mckinnon_data-driven_2024,sizemore_self-consistent_2024,horta_lux_2025} and ab initio (or simulation-based, \citealp[e.g.,][]{rix_constructing_2016,ting_accelerated_2016,ting_payne_2019, obriain_cycle-starnet_2021, straumit_zeta-payne_2022}) methods. Emulation significantly accelerates the process of fitting spectra with fewer of the issues of discriminative classifiers. Because we are modifying the underlying physics of our synthesis, we use a Payne method \citep{ting_payne_2019}, where we train a neural network emulator on synthetic spectra to perform spectral synthesis at a fraction of the time and resources.

Our NNEs are implemented in Python using PyTorch \citep{paszke_pytorch_2019}. We use a simple architecture, similar to the Payne \citep{ting_payne_2019}\footnote{or as in \url{https://github.com/tingyuansen/The_Payne}, which implemented differently than in \citet{ting_payne_2019}.}, as implemented in \citet{kovalev_non-lte_2019} and \citet{storm_observational_2026}: a fully connected, multilayer perceptron. Our NNE has 2 hidden layers with 300 nodes, uses a leaky ReLU activation function between each layer, and passes the final output through a sigmoid function, since the result is a normalized spectrum that should only be allowed to attain values between 0 and 1.

\begin{figure*}
    \centering
    \includegraphics[width=\linewidth]{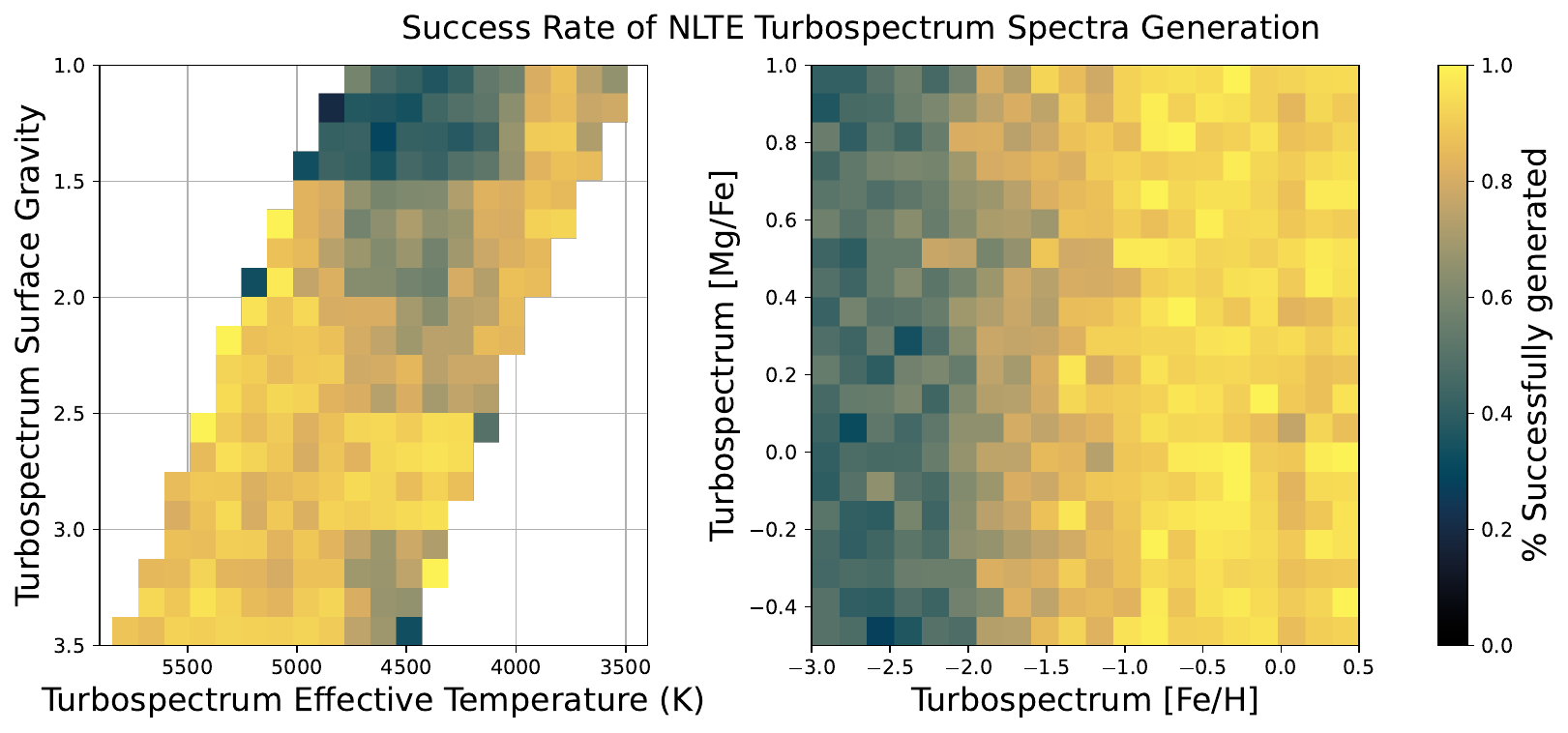}
    \caption{The success rate of NLTE spectra generation by Turbospectrum, over the 20,000 desired spectra for our training. The region around \teff= 4500\,K, \logg= 3.5 comes from missing MARCS models at low metallicities, and accounts for the missing LTE spectra. The other regions where NLTE spectra had trouble generating were from missing NLTE departure coefficients at low metallicities. }
    \label{fig:nlte_success}
\end{figure*}

We train two Paynes: one on LTE syntheses, and one with NLTE for our 8 elements. Our synthetic training grid varies in 15 labels: \teff, \logg, \feh, microturbulence (\vt), \xfe{C}, \xfe{N}, \xfe{O}, \xfe{Na}, \xfe{Mg}, \xfe{Al}, \xfe{Si}, \xfe{Ca}, \xfe{Ti}, \xfe{Mn}, \xfe{Ni}. The ranges for these labels are listed in Table \ref{tab:training_grid} and displayed for \teff, \logg, \alphafe, and \feh in Figure \ref{fig:dr19}. We uniformly sample in all dimensions to create 20,000 sets of stellar labels, from which we generate both LTE and NLTE spectra from 15000 to 17000 {\AA}, at a spacing of 0.1 {\AA}. Not all syntheses were successful, yielding $\sim 18300$ LTE spectra and $\sim 15500$ NLTE spectra. We show in Figure \ref{fig:nlte_success} that the missing spectra are at low metallicities, where APOGEE is already limited, and come primarily from a lack of departure coefficient coverage.

We set aside a random 30\% of the synthetic spectra to use as the validation set, with the remaining 70\% as the training set. At each step of training, we calculate the L1 loss (i.e. the mean of the absolute residuals) of all the spectra in the training set. The parameters are adjusted to minimize the training loss, using an RAdam optimizer \citep{liu_variance_2019} with a constant learning rate of $0.003$, calculated in batches of 512 spectra. Every 100 steps, the L1 loss is calculated from the validation set, and if the validation loss has decreased, the model is saved.

Our training is analogous to the \emph{sparse grid} procedure from \citet{ting_payne_2019}, training on a uniformly sampled grid, rather than their \emph{refined grid} procedure, where \xfe{X} follows the distributions of APOGEE to constrain flux variation. As such, to limit the complexity of our model, we restricted our training to just the red giant branch, as opposed to their model, which could fit both dwarfs and giants. We also trained on $\sim$15,000--18,000 spectra, rather than their 2,000 spectra. Future implementations of the Payne4GAIN will involve experimentation with a refined grid and modified architectures, as part of expanding the parameter range and improving the accuracy of our NNEs \citep[e.g.,][]{storm_observational_2026}. Our implementation of the Payne differs in a few more ways: we use 15 labels with 11 chemical abundances, compared to their 25 labels with 20 abundances, they use \code{ATLAS12} model atmospheres with the \code{SYNTHE} spectral modelling code \citep{kurucz_atlas12_2005,kurucz_atlas12_2013}, and they independently tuned their linelist for their analysis.

\begin{table}[]
    \centering
    \caption{The label ranges of our training grid}

\begin{tabular}{c|c c}
\hline
    Label & Minimum & Maximum \\ \hline
    \logg &  1& 3.5\\
    \teff~$^\dagger$ &  $\sim3500$ & $\sim 5900$\\
    \feh & $-2.5$&0.5\\
    $v_t$ & 1&3\\
    \xfe{C} & $-1.25$&1\\
    \xfe{N} & $-0.5$&1.25\\
    \xfe{O} & $-0.5$&1\\
    \xfe{Na} & $-0.75$&1\\
    \xfe{Mg} & $-0.5$&1\\
    \xfe{Al} & $-0.75$&1\\
    \xfe{Si} & $-0.5$&1\\
    \xfe{Ca} & $-0.5$&1\\
    \xfe{Ti} & $-0.5$&1\\
    \xfe{Mn} & $-0.75$&1\\
    \xfe{Ni} & $-0.5$&1\\
    \hline
\end{tabular}

    \label{tab:training_grid}

    \vspace{0.5ex}
    \footnotesize
    $\dagger$ Range of \teff is a function of \logg, as in Fig. \ref{fig:dr19}.
\end{table}

\subsection{Fitting Spectra}
Once the Payne is trained, it is straightforward to fit observed spectra using chi-squared minimization. Our fits are initialized with the ASPCAP raw values. We fit the spectra in two stages: first, we hold the elemental abundances at solar value and fit the stellar parameters. Then, we fix the stellar parameters and fit the chemical abundances. This two-stage process is analogous to the two stages of ASPCAP fitting, except we do not employ wavelength windows at either stage. We found that this process yielded chemical abundances and stellar parameters that agreed with ASPCAP better than an unconstrained fit.

We apply two masks during fitting: the first is the wavelength mask that is packaged with the \code{mwmStar} files from MWM DR19, and the second is a wavelength mask based on our NNE's agreement to Arcturus: using Arcturus' raw stellar labels from ASPCAP, we generate a spectrum using our LTE Payne, then mask pixels that deviate by more than 0.02 (in normalized flux). For NLTE Payne, we do the same but with the calibrated stellar labels, resulting in a similar mask. This procedure comes from \citet{ting_payne_2019}, and represents imperfections in the linelist or observational problems, like the line spread function or telluric contamination. Similar to their spectroscopic mask, we mask around $\sim 10\%$ of pixels for both LTE and NLTE fitting.

Simultaneously with the NNE, we fit three 2nd order polynomials for the continuum (one for each chip). APOGEE DR19 spectra are already continuum normalized, but sometimes the normalization is imperfect. We also fit a gaussian broadening factor and radial velocity offset (which is close to zero, since the spectra is already at rest-frame). Gaussian broadening is an oversimplification of the line-spread function, which changes as a function of wavelength and fiber \citep{saydjari_improving_2025}. We intend to implement the detailed line-spread function in a future VAC. This means that our model consists of 26 total parameters: 15 labels, 9 continuum parameters, Gaussian broadening, and radial velocity. After generating the model at 0.1{\AA} spacing, we downsample onto the APOGEE wavelength solution to calculate $\chi^2$.

\subsection{Selecting APOGEE Spectra}
\label{subsec:selecting}

From the \code{astraAllStarASPCAP} catalog, we removed entries with \code{flag\_bad}, and those that had invalid values of \code{raw\_teff}, \code{raw\_logg}, \code{raw\_fe\_h}, and \code{raw\_mg\_h}. We then selected all stars (Fig. \ref{fig:dr19}) whose raw values fell within the label ranges of our training grid for  \teff, \logg, \vt, \feh, and \xfe{Mg}. This resulted in $\sim$360,000 spectra to fit with the LTE NNE and the NLTE NNE. Each fitting of the spectra with a single NN was done using 100 cores of the University of Utah CHPC Cluster in less than 48 hours. The fitting results of the $\sim$360,000 stars compose our VAC, which we detail in Section \ref{sec:conclusion}. 

\begin{figure}
    \centering
    \includegraphics[width=\linewidth]{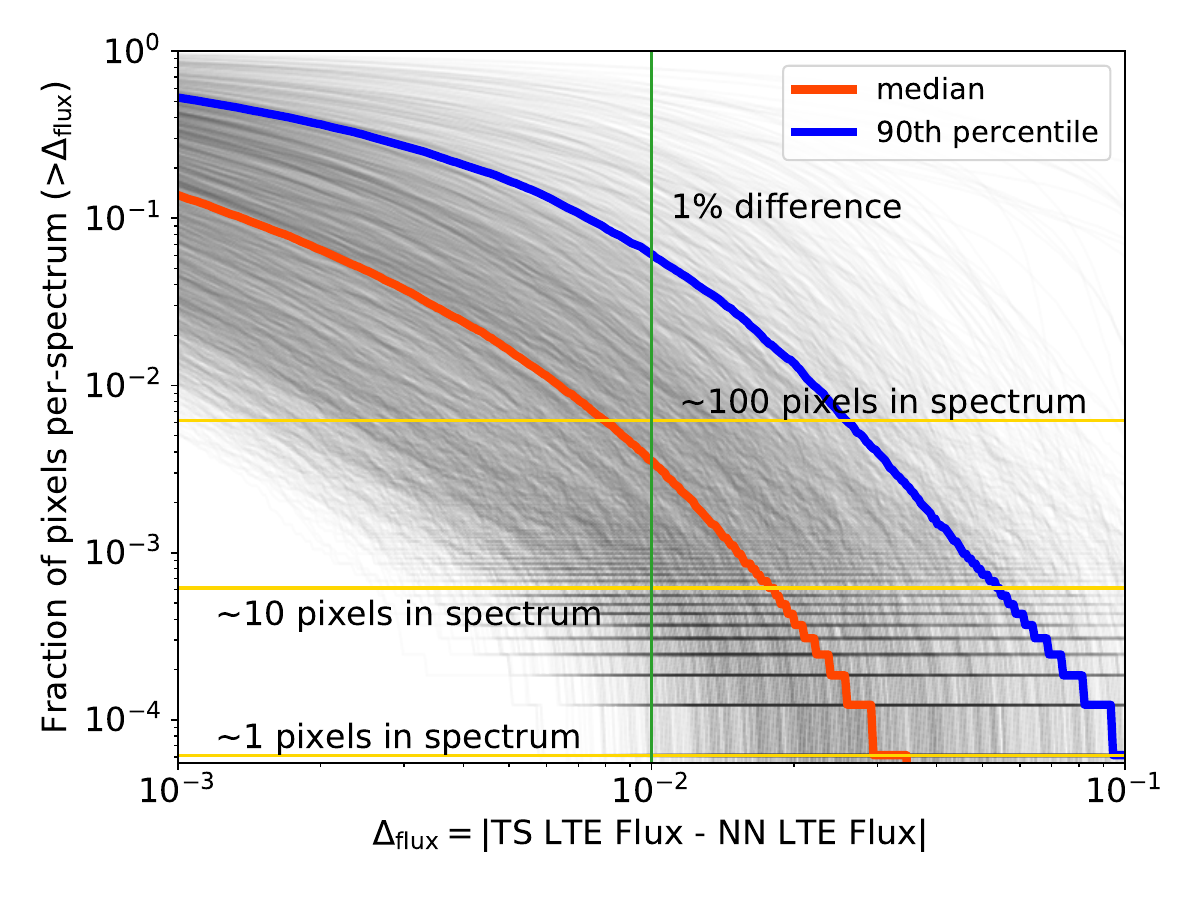}
    \caption{Survival functions of the L1 loss of the LTE NNE output, compared to a set of $\sim 3000$ Turbospectrum-generated LTE spectra at the same labels. Each black curve is the survival function of a single spectrum and the red and blue are the median and 90th percentile survival curves.}
    \label{fig:pixelmatch}
\end{figure}
\begin{figure}
    \centering
    \includegraphics[width=\linewidth]{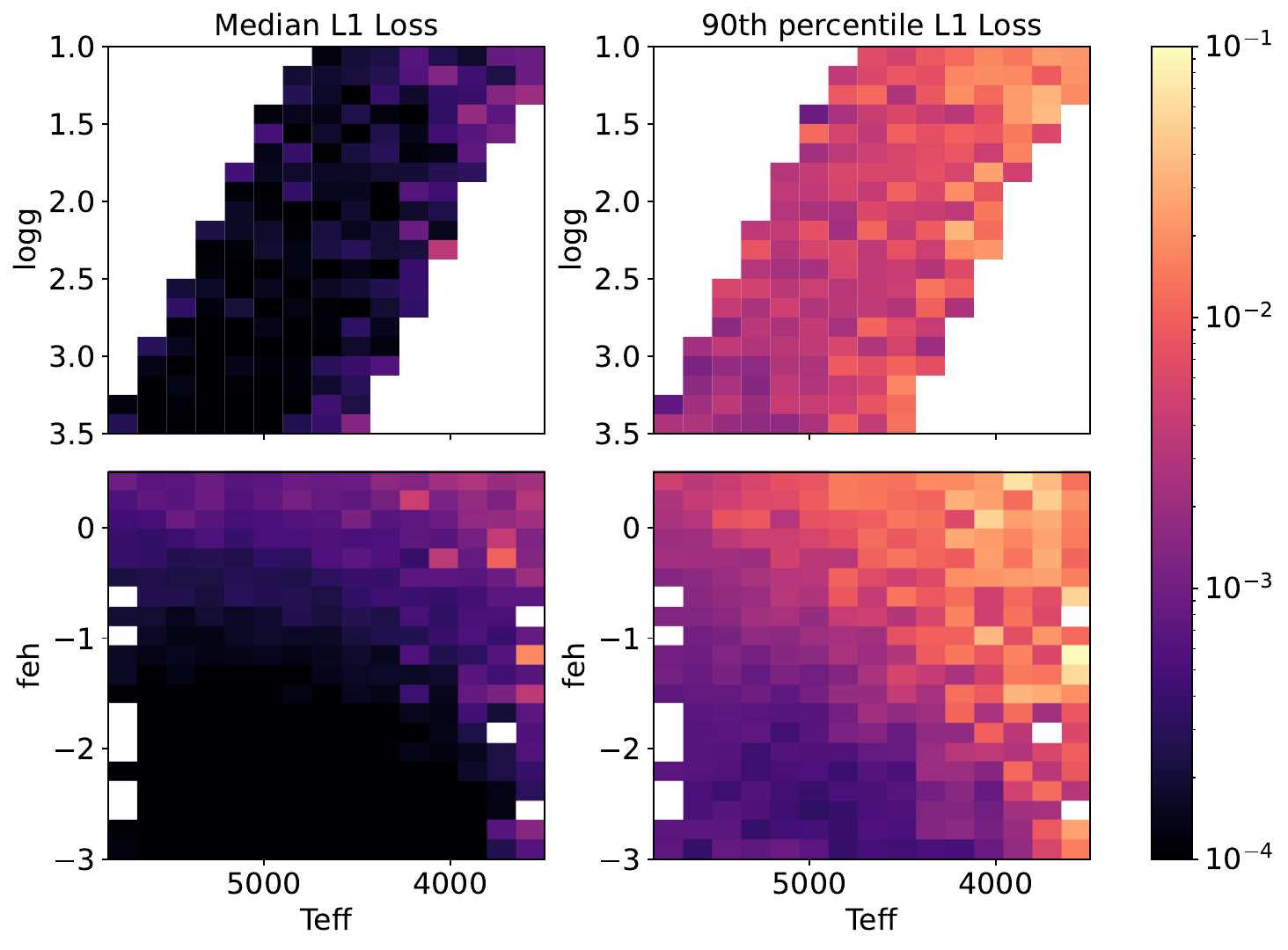}
    \caption{The median and 90th percentile L1 losses the LTE NNE, compared to a set of $\sim 3000$ Turbospectrum-generated LTE spectra at the same labels, presented as a function of the stellar parameters. The losses for the NLTE NNE are similar.}
    \label{fig:pixelkiel}
\end{figure}

\begin{figure*}[h]
    \centering
    \includegraphics[width=\linewidth]{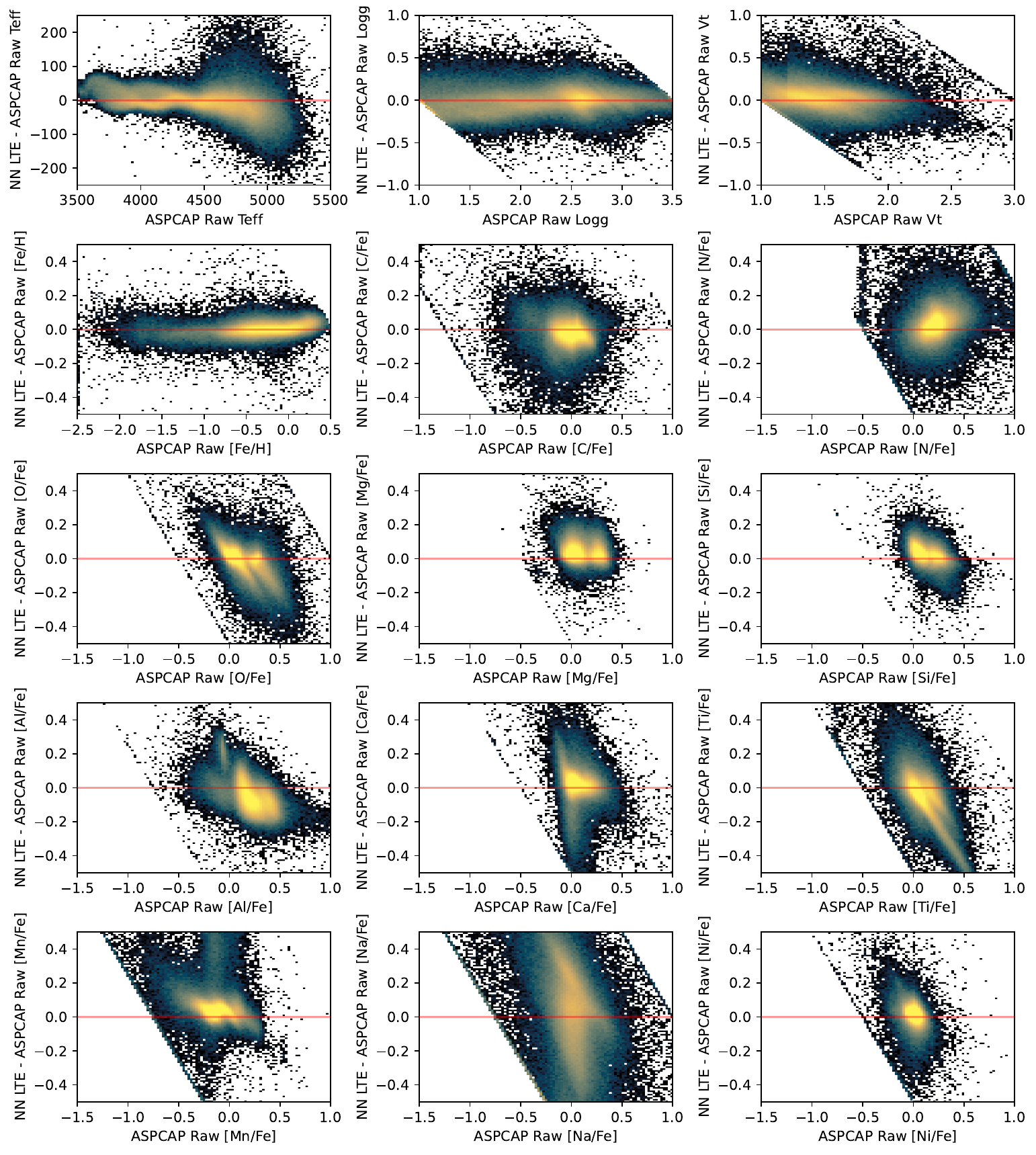}
    \caption{Comparison between the ASPCAP raw parameters and those from our LTE NN fits. There is good agreement for stellar parameters, and general agreement for chemical abundances. Note that the vertical axes of the plots are more zoomed in than the horizontal axes.}
    \label{fig:lte_comparison}
\end{figure*}

\section{Validating the Payne4GAIN}
\label{sec:validation}
We demonstrate that our LTE Payne accurately reproduces both Turbospectrum and results from ASPCAP.

\subsection{Comparing Payne4GAIN LTE to Turbospectrum}
To determine the accuracy of our neural network training, we generate an independent set of $\sim3000$ LTE spectra and compare them to the P4G-generated output at the same input parameters. In Figure \ref{fig:pixelmatch}, we plot the survival functions of this comparison: each line plots the fraction of pixels in each NN-synthesized spectrum (vertical) that differs from the reference spectrum by more than some amount (horizontal). We also plot the median and 90th percentile survival curves. The median NNE output differs from the fiducial Turbospectrum spectra by 0.01 (in normalized flux) for less than 0.5\% of the output pixels generated. With a target SNR of around 100, we expect spectra from MWM to be accurate to around 1\%. For the median case, no pixels differ by more than 0.03. In the 90th percentile case, only around 1\% of pixels differ by more than 0.02 between the Turbospectrum and NN outputs at the same labels. We consider this to be sufficiently accurate for our simple architecture. Improvements could be made at the cost of a larger model, more data, or slower runtime for the emulator (e.g. \citealp{rozanski_scaling_2025}).

To further elaborate on the quality of our neural network, we also track how the L1 loss (or mean absolute error) changes as a function of our stellar parameters. This is displayed in Figure \ref{fig:pixelkiel}. Our neural network becomes less accurate at low temperatures and high metallicities, which is unsurprising because these are the regimes where molecular features become dominant.

\subsection{Comparing Payne4GAIN LTE to ASPCAP}
ASPCAP \citep{garcia_perez_aspcap_2016, holtzman_apogee_2018, jonsson_apogee_2020, meszaros_sdss-v_2025} is the stellar parameter and chemical abundance pipeline for APOGEE.  It fits spectra using \code{FERRE} \citep{allende_prieto_spectroscopic_2006}, which takes large grids of precomputed spectra and interpolates them to find the best stellar labels. \code{FERRE} uses Principal Component Analysis (PCA) during fitting to reduce the dimensionality and memory-usage of their grids. This is still quite resource-intensive, and has the added drawback that it limits the number of dimensions the grid can vary over.

The ASPCAP spectral grid varies in 7D: \teff, \logg, \feh, \vt,\alphafe,\xfe{C},\xfe{N}. First, the stellar parameters are determined through a full-spectrum fit, in these dimensions. Then, the stellar parameters are held fixed, and the elemental abundances are fit in \emph{windows}, using regions of the spectrum that are particularly sensitive to only the given element. Rather than directly varying those elements, \xh{X} is fit by varying \feh at solar \xfe{X} (or by varying \xfe{C} and \xfe{N} for C and N, or \alphafe for O, Mg, Si, S, Ca, and Ti). 

Like ASPCAP, our LTE NNE method uses MARCS model atmospheres and uses the same linelist, so we expect the results to be fairly similar. However, there are a few notable differences. We use Turbospectrum, which is able to perform radiative transfer using spherical geometry, while ASPCAP uses \code{SYNSPEC} \citep{hubeny_brief_2017}, which assumes plane-parallel geometry. Our synthetic grid varies elements independently, affecting the chemical balance used to calculate molecular abundances. This is relevant because we do not use the wavelength windows of ASPCAP, so molecular lines play a role in our fits. There is also a difference that ASPCAP's model atmospheres vary in \alphafe, \xfe{C}, and \xfe{N}, whereas our model atmospheres don't vary those parameters, only our syntheses; however, we checked that the differences this produces in the APOGEE DR17 spectral grid \citep{abdurrouf_seventeenth_2022} are small.

In Figure \ref{fig:lte_comparison}, we compare the residuals of the labels of our LTE NNE fits to those from ASPCAP, as a function of those labels. We see very good agreement for the stellar parameters, \teff, \logg, \vt, and \feh. We see broad agreement for the abundances, although we note that the Na fit is very poor, and the Ti fit has a strong artifact. For sodium, determination of abundance depends on two fairly weak lines, and this element is characterized as being ``low-precision" in ASPCAP as well \citep{meszaros_sdss-v_2025}. One of the Na lines also coincides with a Diffuse Interstellar Band, further complicating accurate modelling \citep{mckinnon_data-driven_2024}. 

For Ti, this seems to come from an inability of fitting high-metallicity stars (\feh $>0$). Artifacts in \xfe{O},\xfe{Mn}, and \xfe{Ca} can also be attributed to poor fits for high-metallicity stars, suggesting that these poor fits are from complications involving molecular or blended features. However, these features are also present in comparisons of the NLTE NNE abundances with ASPCAP. While agreement with ASPCAP is relevant in considering the performance of the neural networks, we also emphasize that the primary difference between our NLTE and LTE neural networks are the presence or absence of NLTE physics going into the line calculations, which is why we present NLTE corrections to the ASPCAP abundances, in addition to our direct NNE fits.

As seen in Figures \ref{fig:direct_feh_0} and \ref{fig:direct_feh_1}, the \xfe{X} trends from raw ASPCAP and P4G LTE result in generally similar trends, with the biggest differences for Al, Ti, and Mn. As we explain in Section \ref{subsec:Ti}, the discrepancy in Ti can be explained by the different behavior of Ti I and Ti II lines, and the fact that our method fits the entire spectrum. The full-spectrum fitting is also the strongest explanation for the differences in Al and Mn.

\section{Results}
\label{sec:differences}

\subsection{Raw ASPCAP vs LTE NNE vs NLTE NNE}
\begin{figure*}
    \centering
    \includegraphics[width=\linewidth]{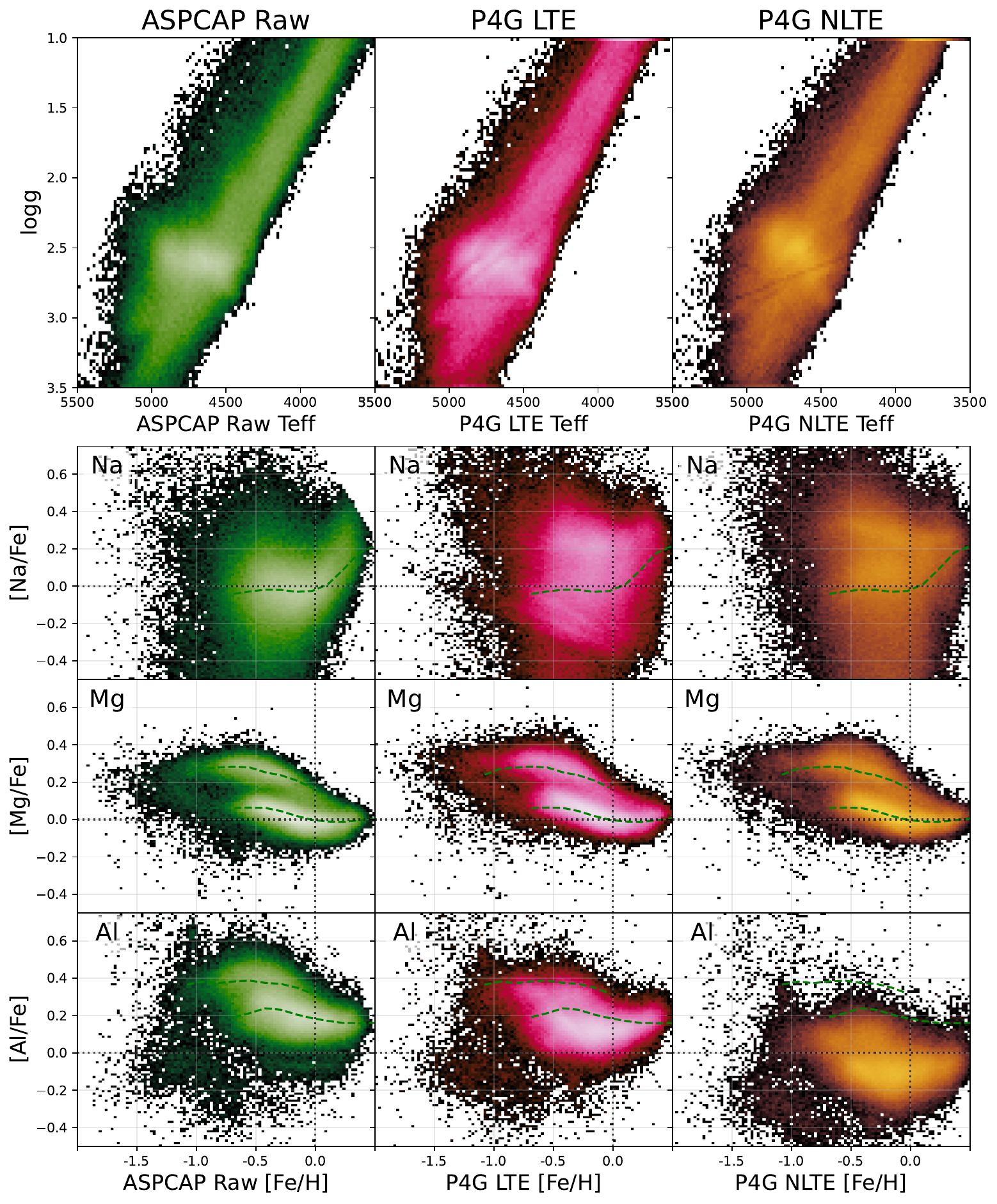}
    \caption{Kiel Diagrams and trends of \xfe{X} vs \feh for Na, Mg, Al with parameters and abundances from ASPCAP (raw), P4G LTE and P4G NLTE. The lines in the P4G Kiel Diagrams are associated with discontinuities in our NNEs, but do not greatly impact the fitting of abundances. Note that in the P4G NLTE, \feh is in LTE. The black dotted lines show solar \feh and \xfe{X}. The green dashed lines show the median ASPCAP raw \xfe{X} trends, for comparison with the other methods.}
    \label{fig:direct_feh_0}
\end{figure*}

In the top panel of Figure \ref{fig:direct_feh_0}, we show the Kiel diagram for each of the ASPCAP raw, Payne4GAIN LTE, and Payne4GAIN NLTE parameters. The P4G LTE stellar parameters show good agreement to those from ASPCAP, as previously demonstrated. The P4G NLTE parameters display a notable \logg offset ($\sim -0.1$\,dex), which brings it into better agreement with ASPCAP surface gravities ($-0.09$\,dex for RGB stars, $-0.18$\,dex for red clump stars, \citealp{meszaros_sdss-v_2025}). ASPCAP \logg is calibrated on APOKASC asteroseismic surface gravities \citep{pinsonneault_apokasc-3_2025}.

In the P4G Kiel diagrams, we can see some line-shaped underdensities. The source of these artifacts is not known, and does not correspond to differences in fit quality (as measured by $\chi^2$). Stars whose parameters occupy these artifacts also do not have systematically different abundances than those to the sides of the artifacts. Still, these artifacts are one of the reasons we advocate for the use of our NLTE corrections, rather than the direct labels fit by the NNEs. These artifacts will likely be mitigated with more complex NNE training and fitting.

\begin{figure*}
    \centering
    \includegraphics[width=\linewidth]{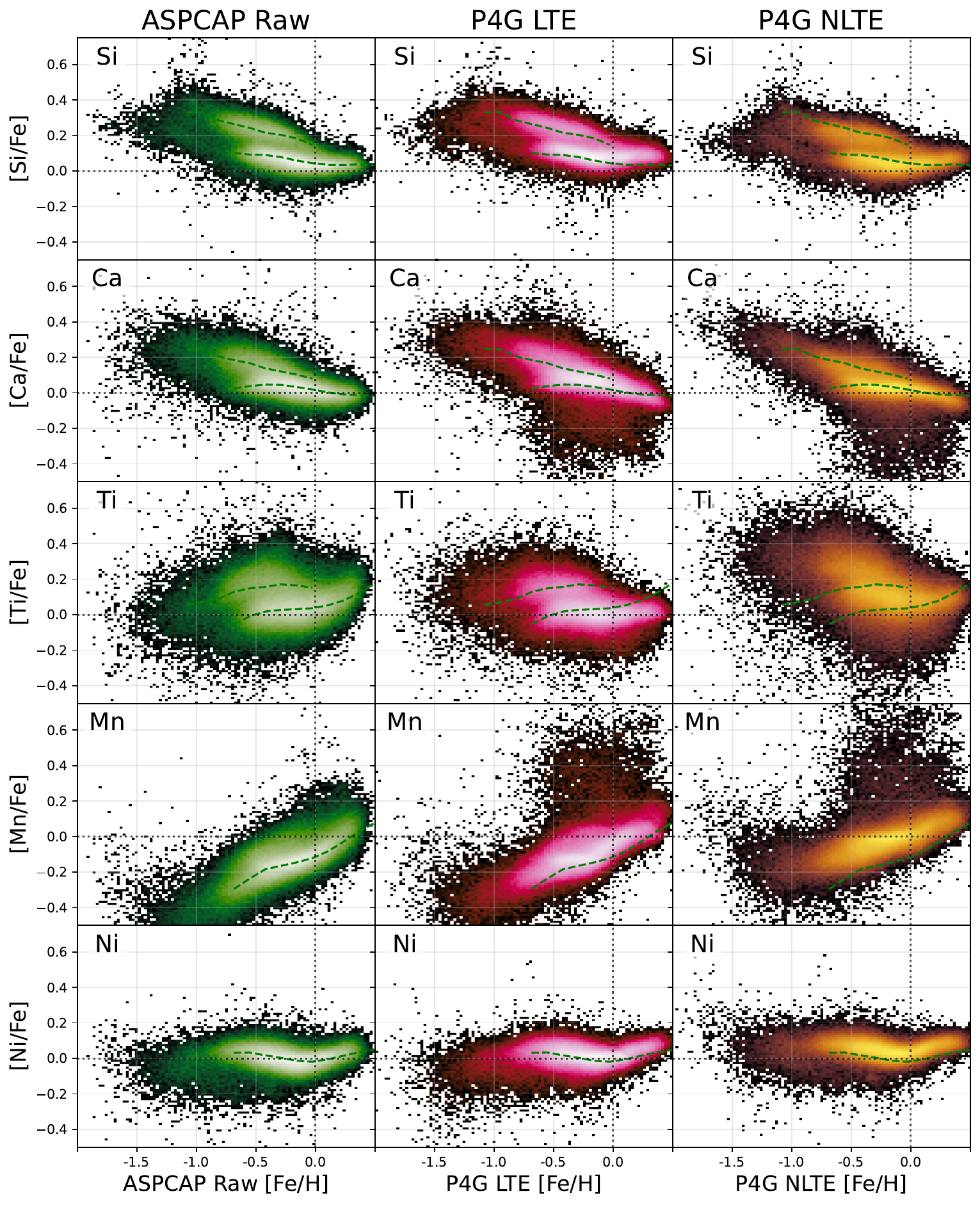}
    \caption{same as Figure \ref{fig:direct_feh_0}, but \xfe{X} for Si, Ca, Ti, Mn, Ni}
    \label{fig:direct_feh_1}
\end{figure*}

We show in Figures \ref{fig:direct_feh_0} and \ref{fig:direct_feh_1} the chemical abundance trends of \xfe{X} vs \feh for the three methods: ASPCAP raw, P4G LTE, and P4G NLTE. As previously stated, \feh is calculated in LTE for all three methods. We discuss these trends on an element-by-element basis in Section \ref{sec:discussion}. Overall, we find that the biggest NLTE effects occur for the elements Al, Ti, and Mn, with lesser effects for Si and Ni. 

We note that the P4G NLTE trends are more diffuse than the LTE trends, which we attribute to the LTE-tuned linelist. This, in addition to the systematics of the NNE procedure, motivate us to present our results as NLTE-LTE corrections. 

\subsection{NLTE-LTE corrections}
\label{subsec:corr}
\begin{figure*}
    \centering
    \includegraphics[width=\linewidth]{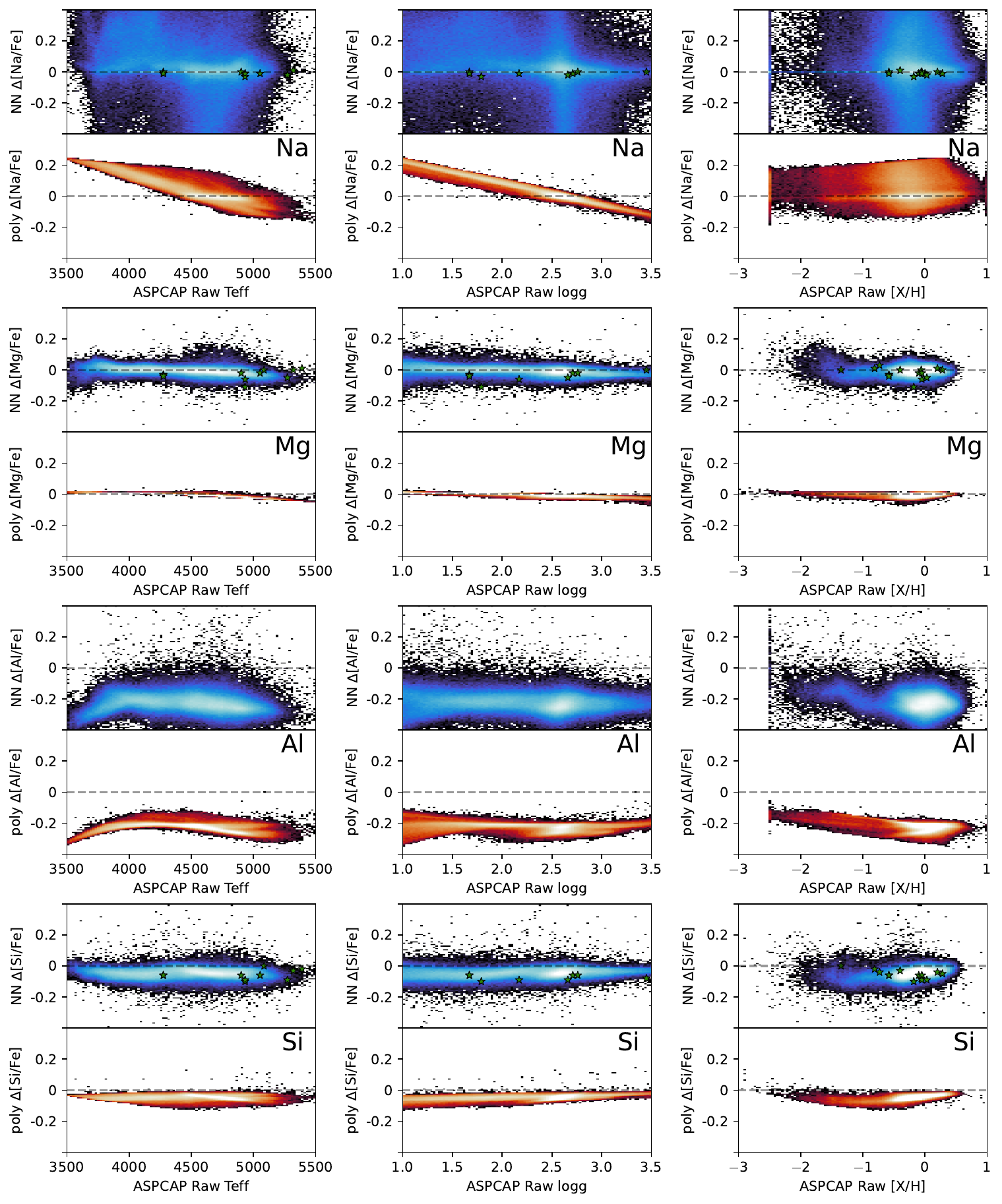}
    \caption{NLTE differences for \xfe{Na} (first two rows of subplots), \xfe{Mg} (third, fourth rows), \xfe{Al} (fifth, sixth rows), and \xfe{Si} (last two rows). The top panels of each pair, with the blue histogram, are the per-star NLTE differences that come from fitting each star with each Payne. The latter panels of each pair, with the orange histogram, are the polynomial NLTE differences that come from fitting the per-star NLTE differences. The corrections for \xfe{X} are presented in dex against \teff (first column), \logg (second column), and \xh{X} (third column), the variables in the polynomial corrections (Table \ref{tab:equations}).  The green stars correspond to the corrections from the stellar sample of \citet{zhou_nlte_2023,zhang_nlte_2017,zhang_nlte_2016}.}
    \label{fig:nlte_corrs_0}
\end{figure*}

\begin{figure*}
    \centering
    \includegraphics[width=\linewidth]{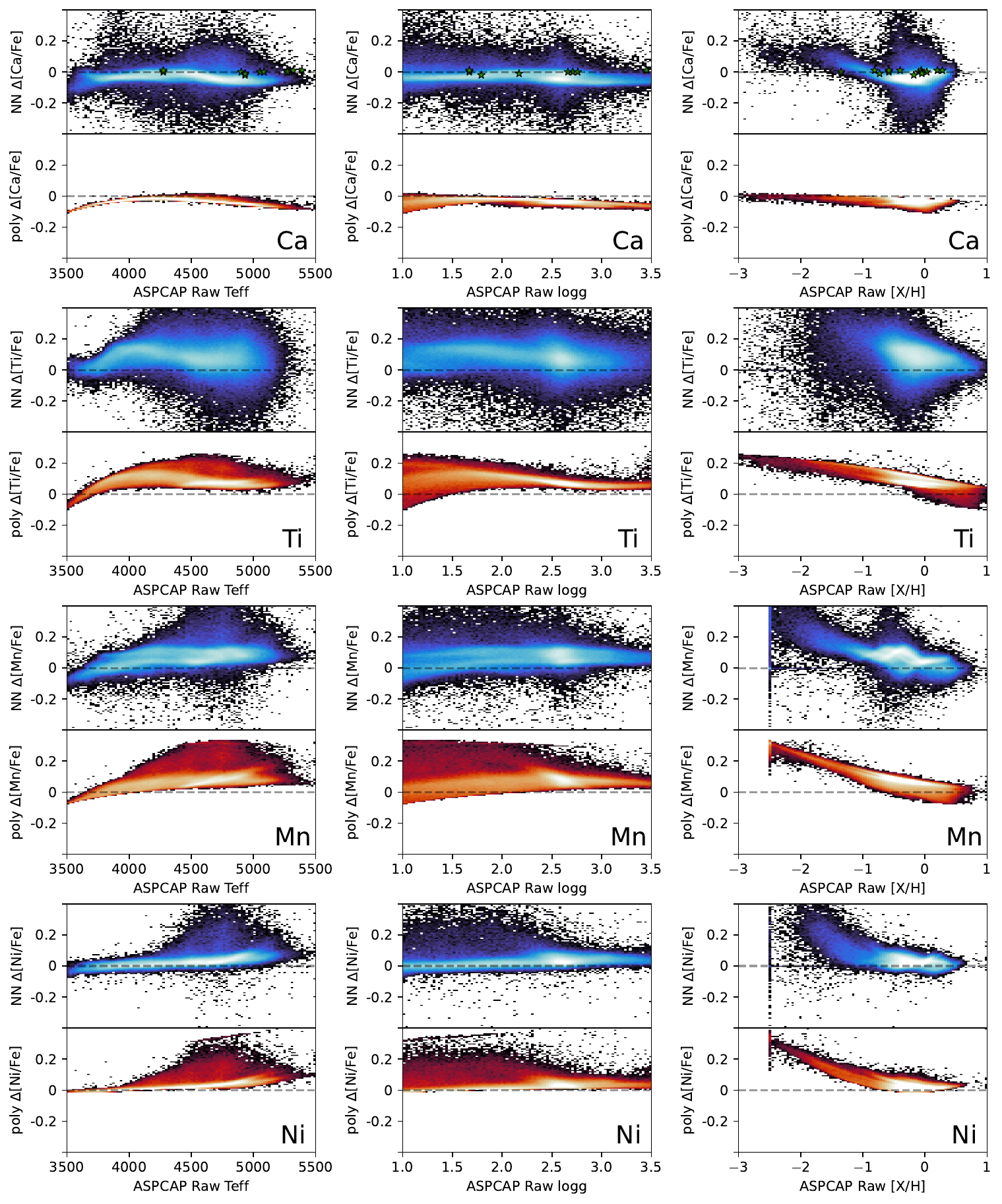}
    \caption{The same as Figure \ref{fig:nlte_corrs_0}, but for \xfe{Ca} (first two rows), \xfe{Ti} (third, fourth rows), \xfe{Mn} (fifth, sixth rows), and \xfe{Ni} (last two rows). The green stars correspond to the corrections from the stellar sample of \citet{zhou_non-lte_2019}.}
    \label{fig:nlte_corrs_1}
\end{figure*}

Fitting each spectrum with the NLTE NNE and the LTE NNE, we can take the difference of the labels to get NLTE-LTE differences for each star. This is plotted in the top panels in the subplots of Figures \ref{fig:nlte_corrs_0} and \ref{fig:nlte_corrs_1} (the blue histograms). 

We expect NLTE effects to vary smoothly as a function of stellar parameters. Deviations from this behavior mostly arise from data noise or the accuracy of the NNE. As such, we fit the NLTE corrections to polynomial functions of \teff, \logg, and \xh{X} (which encompass \feh) for each element. The order of the polynomials were hand-selected to be small enough while capturing the parameter-dependent behavior of the NLTE corrections, up to an order of 3. These are presented in the lower panels in the subplots of Figures \ref{fig:nlte_corrs_0} and \ref{fig:nlte_corrs_1} (the orange histograms). The equations of those polynomial fits are in Table \ref{tab:equations}. In most cases, the behavior of the corrections below \xh{X}$\sim -2$ is an extrapolation of the trend of the correction, due to a lack of stars that are both metal-poor and element-poor.

We apply these polynomial-smoothed corrections to the ASPCAP raw abundances to get \emph{P4G-corrected ASPCAP abundances}. These combine the advantages of ASPCAP's better treatment of continuum and line spread functions with the systematic effects of NLTE from our analysis. 


\begin{table*}
    \centering
    \caption{Polynomial Corrections for NLTE, based off of stellar parameters}
    \begin{tabular}{c|cccccccccc}
[X/Fe] & 1 & $T$ & $T^2$ & $T^3$ & $L$ & $L^2$ & $L^3$ & $X$ & $X^2$ & $X^3$\\
\hline
Na & 0.13 & -0.048 &  &  & -0.11 &  &  & 0.0077 &  & \\
Mg & -0.00015 & -0.026 & -0.016 &  & -0.0049 & 0.00067 &  & 0.0018 & 0.0013 & \\
Al & -0.25 & -0.15 & -0.089 & 0.13 & 0.016 & 0.027 &  & -0.028 &  & \\
Si & -0.1 & -0.083 & 0.024 &  & 0.072 & -0.0066 &  & -0.0049 & 0.0039 & \\
Ca & -0.026 & -0.047 & -0.064 & 0.056 & 0.0016 & -0.0013 &  & -0.0081 &  & \\
Ti & 0.12 & -0.049 & -0.071 & 0.15 & -0.023 & -0.029 & 0.015 & -0.024 &  & \\
Mn & 0.089 & 0.092 & -0.068 &  & -0.026 & -0.0059 &  & -0.016 & 0.0055 & \\
Ni & 0.025 & 0.066 & 0.047 &  & 0.013 & -0.016 &  & 0.0058 & 0.013 & \\
\end{tabular}
    
    \vspace{5pt}
    Here, $T$=\teff/1000$-4.5$, $L$=\logg$-1.5$, $X$=\xfe{X}/0.5
    \label{tab:equations}
\end{table*}

\subsection{Abundance Zero Points}
Solar-metallicity solar neighborhood (SMSN) stars have been shown to have \xfe{X} very similar to the Sun across a range of stellar parameters \citep{reddy_chemical_2003, adibekyan_chemical_2012, bensby_exploring_2014}. Thus, in MWM DR19, the only calibration that is applied to the abundances (except C and N) of all stars is a zero-point offset that forces a SMSN sample to have mean \xh{X}=0. This is generally a small adjustment to correct systematic discrepancies from stellar modeling (e.g. \citealp{jofre_accuracy_2019}). The callibration is applied separately for giants and dwarfs, and we display the giant offsets from \citet{meszaros_sdss-v_2025} in the first column of Table \ref{tab:zeropoint}. Of our 8 elements, Al and Mn have calibrations larger than 0.1\,dex, with the rest smaller than $\sim 0.05$\,dex.

To calibrate the abundances in our analysis, we select a similar SMSN sample, restricting \code{raw\_m\_h\_atm} to within $\pm 0.05$dex, \code{plx} to be larger than 2\,mas (distance smaller than 500\,pc), \code{snr} greater than 50, and no \code{flag\_bad}. Cross-matching with our catalog, we get 332 SMSN stars that were fit with the Payne4GAIN. The mean abundances (P4G LTE, P4G NLTE, and the P4G-corrected raw ASPCAP) for these P4G SMSN stars compose the latter three rows of Table \ref{tab:zeropoint}.

The P4G LTE zero-point offsets are generally small, with only Na and Al having offsets larger than 0.1\,dex. Notably, the P4G Ti offset is smaller than that of ASPCAP. The P4G NLTE zero-point offsets are smaller than the P4G LTE offsets for Na, Mg, Si, Ca, and Mn, but not for Al, Ti, or Ni. The zero-point offsets for the P4G-corrected raw ASPCAP values are smaller in magnitude than those from MWM DR19 \citep{meszaros_sdss-v_2025} for Al, Si, Ca, and Mn, but not for Na, Mg, Ti, and Ni. For Na and Ni, these offsets are similar to those reported in MWM DR19. For Mg, the zero-point is still quite small, $<0.05$\,dex, but for Ti, the zero-point is close to 0.1. 

We emphasize that these SNSM zero-point calibrations are not applied to the \code{aspcap\_p4gcorr\_xfe}s in our VAC. For agreement to the APOGEE abundance scale, one could subtract the values of the final column of Table \ref{tab:zeropoint} from the P4G-corrected values. As the zero-point calibrations are not directly caused by the inclusion NLTE physics, we compare and discuss raw spectroscopic abundances in Figures \ref{fig:direct_feh_0} and \ref{fig:direct_feh_1} and in the individual subsections of Section \ref{sec:discussion}.

\begin{table*}[]
    \centering
    \caption{Solar-Metallicity Solar Neighborhood Zero-point Offsets}
    
\begin{tabular}{c|cccc}
         Element& MWM DR19 (Giants) & P4G LTE & P4G NLTE & P4G-Corrected\\
         \hline
Na & $-0.0507$ & $ 0.1520$ & $ 0.0878$ & $-0.05355$ \\
Mg & $-0.0005$ & $ 0.0094$ & $-0.0090$ & $-0.03034$ \\
Al & $ 0.1751$ & $ 0.1090$ & $-0.1090$ & $-0.04737$ \\
Si & $ 0.0234$ & $ 0.0629$ & $ 0.0300$ & $-0.00314$ \\
Ca & $ 0.0323$ & $ 0.0513$ & $ 0.0157$ & $-0.02522$ \\
Ti & $ 0.0470$ & $ 0.0064$ & $ 0.0670$ & $0.09545$ \\
Mn & $-0.1338$ & $-0.0613$ & $-0.0001$ & $-0.05006$ \\
Ni & $-0.0127$ & $ 0.0100$ & $ 0.0354$ & $0.01683$ \\
    \end{tabular}
    
    \label{tab:zeropoint}
\end{table*}

\subsection{Temperature Dependence}
Because a stellar population can occupy a range of \logg and \teff as different mass stars evolve along the same isochrone, we do not expect to see chemical abundance trends with most parameters (aside from \xfe{C} and \xfe{N}, due to CN cycling in RGB stars, \citealp{iben_stellar_1967}). However, we do see spurious dependencies on \teff and \logg for abundances from ASPCAP and some analyses have applied calibrations to correct for these \citep{weinberg_chemical_2022,griffith_similarity_2021, sit_chemical_2024, meszaros_sdss-v_2025}. These analyses have suggested that this could be because ASPCAP doesn't consider NLTE effects, which do have \teff and \logg dependencies. In DR19, while they are not applied to the ASPCAP results, \teff corrections are calculated from open clusters due to the expectation that each cluster represents a chemically identical population \citep{bovy_slar_2016,poovelil_vizier_2022,sinha_comprehensive_2024}.

Thus, to determine if our NLTE corrections alleviate these temperature dependencies, we assemble a sample of open cluster stars by cross-referencing our VAC with the OCCAM catalog \citep{otto_open_2026}. We selected only clusters with more than 6 members within our VAC, that had a \feh spread of less than 0.18, but \teff range of more than 1200\,K. This ensures that each cluster is chemically homogeneous while having enough \teff variation to probe abundance dependencies. This results in 595 stars.

\begin{figure*}
    \centering
    \includegraphics[width=\linewidth]{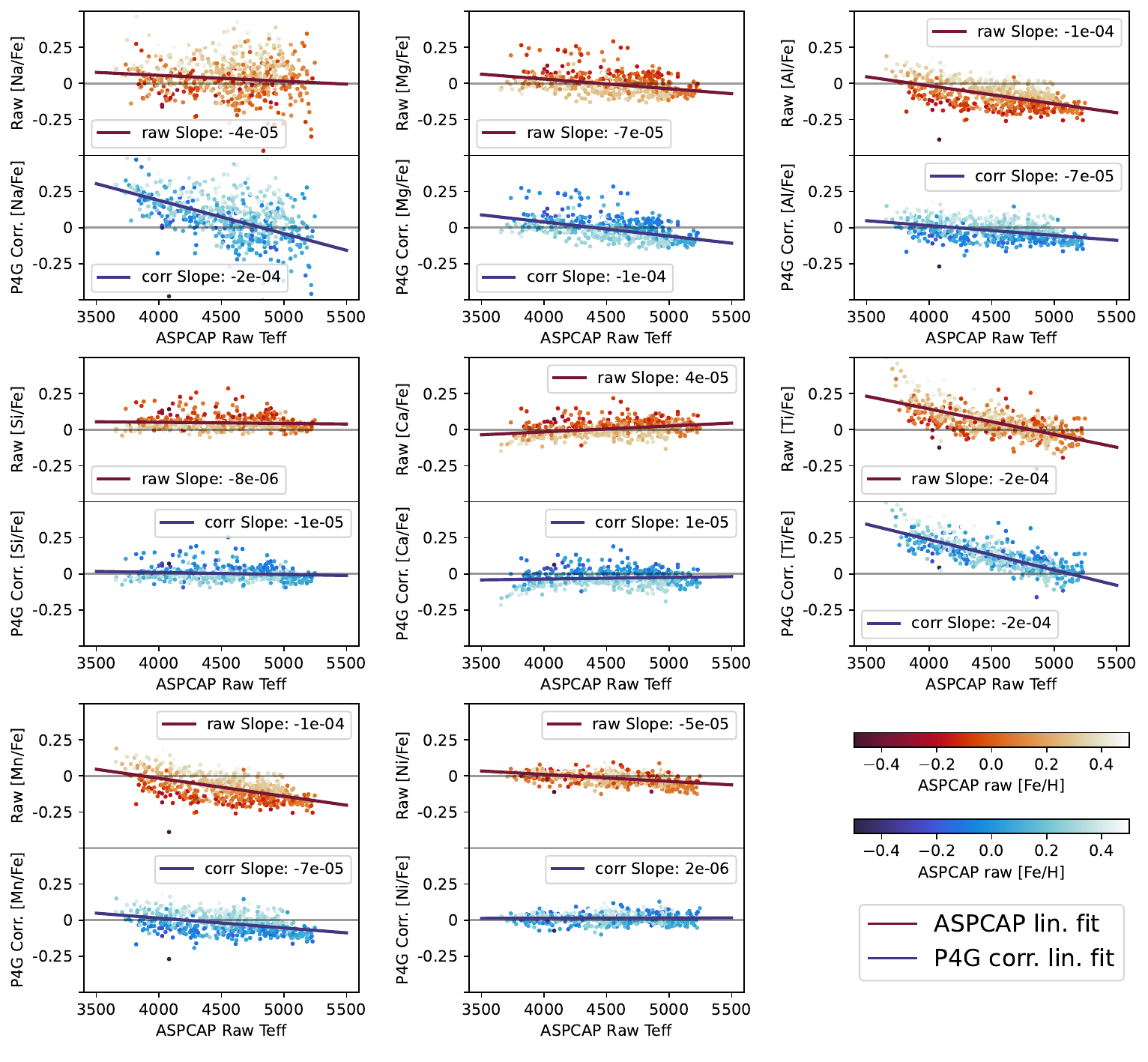}
    \caption{Chemical abundance trends with \teff for 595 OCCAM stars in our VAC. We selected only clusters with a large \teff spread and a narrow \feh spread. We show how the trends lessen, or stay the same, with the addition of our P4G NLTE correction.}
    \label{fig:occam_teff}
\end{figure*}

For the NLTE elements in our analysis, we plot the raw and corrected chemical abundance trends in the OCCAM RGBs in Figure \ref{fig:occam_teff}. We can see for Al, Ca, Mn, and Ni, applying our NLTE corrections lessens the strength of the \teff dependence. For the other elements, Mg, Si, and Ti, the NLTE corrections do not lessen the dependence of the abundance on \teff. We discuss these trends on an element-by-element basis in Section \ref{sec:discussion}.

\subsection{Comparison to other works}

There is a sample of about 12 stars that have been analyzed with LTE and including NLTE for Si \citep{zhang_nlte_2016}, Mg \citep{zhang_nlte_2017}, Ca \citep{zhou_non-lte_2019}, and Na \citep{zhou_nlte_2023}. In their analyses, they derive line-by-line abundances via synthesis to hand-picked lines for each element. As a result, we expect their abundances to be precise, though could differ from our analysis due to differences in method (synthesis, fitting) or in physics assumptions going into NLTE. We plot NLTE-LTE corrections derived from their analyses in Figures \ref{fig:direct_feh_1}  and \ref{fig:nlte_corrs_0}, finding them to be in generally good agreement with the corrections we derive from our Payne4GAIN procedure. These are discussed per-element in Section \ref{sec:discussion}.

\begin{figure}
    \centering
    \includegraphics[width=\linewidth]{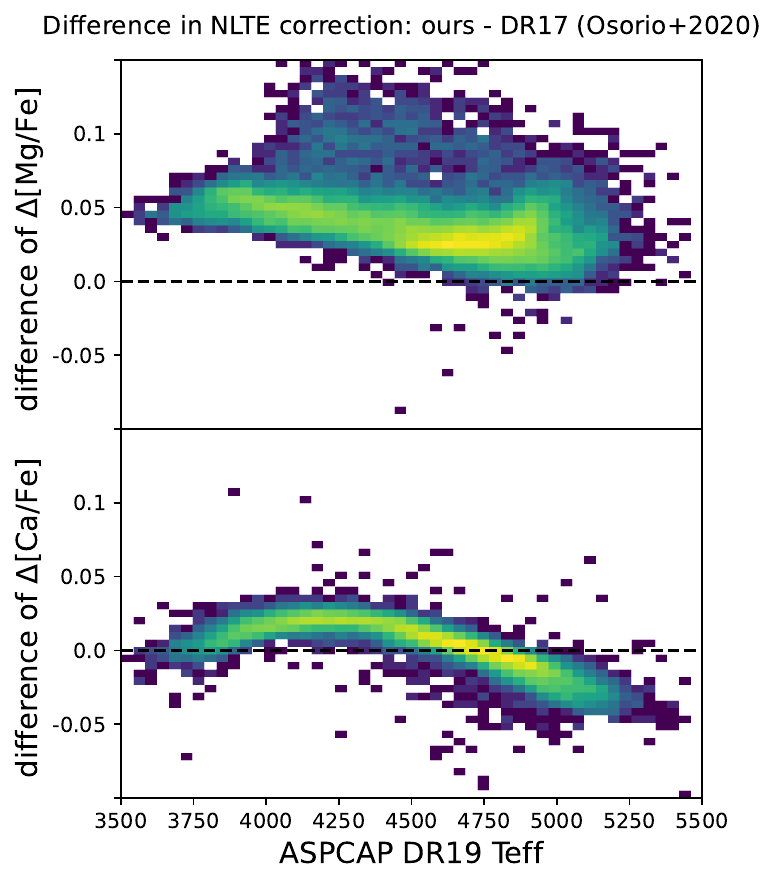}
    \caption{The difference between our NLTE corrections and those of \citet{osorio_nlte_2020}, for Mg and Ca.}
    \label{fig:osorio}
\end{figure}

\citet{osorio_nlte_2020} also conduct an NLTE analysis which resulted in the inclusion of NLTE abundances for Mg, Ca, Na, and K in DR17 \citep{abdurrouf_seventeenth_2022}. Because LTE abundances were also published for DR17, we can compare their NLTE corrections for Mg and Ca to those that we derive for the members of our VAC, which we show in Figure \ref{fig:osorio}. We discuss the differences between our corrections in Sections \ref{subsec:Mg} (Mg) and \ref{subset:Ca} (Ca).

We also qualitatively compare our results to those from \citet{amarsi_galah_2020}, who derived NLTE \xfe{X} trends for several elements in GALAH, an optical spectroscopic survey. We do not necessarily expect different wavelength regimes to have equally strong NLTE differences, but we do expect NLTE \xfe{X} trends to agree between surveys. We discuss the optical NLTE \xfe{X} trends of Na, Mg, Al, Si, Ca, and Mn in Section \ref{sec:discussion}.

\section{Discussion of Individual Elements}
\label{sec:discussion}
\subsection{Sodium (Na)}

Sodium is an odd-Z light element, produced during carbon burning and the NeNa cycle \citep{samland_modeling_1998}. Its final abundance is sensitive to the neutron excess \citep{woosley_evolution_1995}, and by extension, the mass and metallicity of the progenitor star \citep{smiljanic_gaia-eso_2016}.

Unfortunately, Na is one of the least precise elements in ASPCAP because it is measured from two weak lines and can suffer from telluric contamination \citep{meszaros_sdss-v_2025} and diffuse interstellar bands \citep{mckinnon_data-driven_2024}; this is also the case for our analysis, and manifests as a very large scatter, as can be seen in Figure \ref{fig:direct_feh_0}. This leads to a stong slope in the polynomial NLTE corrections for Na that is not accurate. As such, we do not recommend the use of the NLTE corrections we calculate for Na. 

We do, however, point out that \citet{zhou_nlte_2023} calculated LTE and NLTE Na abundances for a handful of stars using spectra from DR17. Corrections calculated from their abundances are plotted as green stars in the first row of Figure \ref{fig:nlte_corrs_0}. These agree well with the NLTE correction for Na from \citet{osorio_nlte_2020} and DR17 \citep{abdurrouf_seventeenth_2022}, which is $\sim -0.02$ for most stars. Trying to interpret the corrections in Na corrections in Figure \ref{fig:nlte_corrs_0}, it would seem that our method results in a slight majority of stars with a very small or zero NLTE correction. This is consistent to the other analyses. Thus, NLTE appears to have a smaller effect for infrared Na abundances than it does for optical Na abundances, in \citet{amarsi_galah_2020}.

\subsection{Magnesium (Mg)}
\label{subsec:Mg}
Magnesium is an $\alpha$-element, and originates almost purely from core collapse supernovae (CCSNe) as opposed to Fe, which is produced by both Type Ia SNe and CCSNe \citep{mcwilliam_abundance_1997,nomoto_nucleosynthesis_2013}. $\alpha$-elements serve as a diagnostic for star formation, with a high-\alphafe plateau indicating enrichment by prompt CCSNe, then a \emph{knee}, followed by a decline due to delayed SNIa enrichment \citep[e.g.][]{mcwilliam_abundance_1997, weinberg_chemical_2019}. 

In the Milky Way, there is a clear bimodality, forming two chemically and kinematically distinct sequences we call the high-$\alpha$ and low-$\alpha$ tracks \citep[e.g.][]{fuhrmann_nearby_1998,prochaska_galactic_2000,bensby_elemental_2003,hayden_chemical_2015,imig_tale_2023}.
Whether the high-$\alpha$ plateau shows trends with metallicity is an open question, one that could be related to the metallicity dependence of CCSNe yields or fast channels of SNIa production \citep{bergemann_non-local_2017,matteucci_new_2006}.

We do not see strong NLTE differences for Magnesium in Figure \ref{fig:direct_feh_0}. The low-$\alpha$ track of the P4G LTE trend appears to rise faster with decreasing metallicity, however, this is also present in the P4G NLTE trend. NLTE also does not reduce the temperature dependence of \xfe{Mg}, as can be seen in Figure \ref{fig:occam_teff}.  Our P4G fits (not pictured) have similar trends to ASPCAP, so this could come from something common to both methods, like a lack of 3D modeling or the tuning of the linelist.

We compare our NLTE corrections to those of \citet{zhang_nlte_2017} in Fig~\ref{fig:nlte_corrs_0}, whose Mg corrections generally agree with ours, but appear systematically lower by $<0.05$ dex. This is consistent with Figure \ref{fig:osorio}, where we show that our NLTE corrections are systematically higher than those of \citet{osorio_nlte_2020}, by around $\sim 0.05$ to $0.02$ dex. \citet{osorio_nlte_2020} attributed differences between their work and \citet{zhang_nlte_2017} to different collision data and also note that \code{MULTI} and \code{TLUSTY} \citep{hubeny_brief_2017} calculate background opacities in different manners---these could explain the discrepancies in our work as well.

NLTE has a smaller effect on infrared Mg abundances than for optical Mg. However, \citet{amarsi_galah_2020} find that the effect of applying NLTE is a flattening of the high-$\alpha$ trend at low metallicities (as opposed to a decline at lower metallicities). This is consistent with the relatively flat trend found in all three IR \xfe{Mg} trends.

\subsection{Aluminum (Al)}

Aluminum, like sodium, is an odd-Z light element,  and is produced during carbon and neon burning during the MgAl cycle \citep{samland_modeling_1998}. Because of the metallicity dependence of their yields, low Na and Al have been used to identify stars formed in low-mass environments, from accreted systems \citep{nissen_two_2010, hawkins_using_2015, das_ages_2020,limberg_reconstructing_2022,horta_chemical_2023,belokurov_dawn_2022}. Interestingly, galactic chemical evolution models have struggled to explain Al and Na at the same time \citep{samland_modeling_1998,smiljanic_gaia-eso_2016}. Al is much better characterized in ASPCAP than Na.

In Figure \ref{fig:direct_feh_0}, our P4G LTE trend is similar to that of the ASPCAP raw \xfe{Al}, though there is a notable zero-point offset for the high-$\alpha$ population of the disk. The ASPCAP raw \xfe{Al} appears to decrease with increasing metallicity, whereas P4G \xfe{Al} increases at super-solar metallicity. It is not clear at the moment which is more accurate, but we note that an increasing trend at high [Fe/H] might be expected since Al is metallicity dependent \citep[e.g.,][with Na]{mcwilliam_chemical_2016}. If so, this may indicate the ASPCAP aluminum windows are blended with other metal lines for cooler, higher-metallicity stars.

Al exhibits a strong difference between LTE and NLTE. Most obviously, there is a clear zero-point offset between LTE and NLTE, as can be seen in our correction, in Figure \ref{fig:nlte_corrs_0}. This results in NLTE corrections between $-0.2$ and $-0.3$ for Al. In DR19, Al is calibrated $0.17$\,dex in the same direction as our NLTE correction, one of the largest in DR19 \citep{meszaros_sdss-v_2025}. It is likely that lack of NLTE contributes strongly to that zero-point offset. Furthermore, NLTE reduces the slope of \xfe{Al} with \teff (Figure \ref{fig:occam_teff}).

NLTE has a larger effect on infrared Al abundances than for optical Al. \citet{amarsi_galah_2020} find NLTE differences in red giants for Al to be around $-0.05$\,dex, though they are in the same direction as our infrared NLTE corrections. Our P4G LTE and NLTE trends rising towards supersolar metallicities also matches better with their \xfe{Al} trends than ASPCAP does.

\subsection{Silicon (Si)}

Silicon is an $\alpha$-element, like Mg, and originates mostly from CCSNe, though has non-negligible contribution from SNIa \citep{mcwilliam_abundance_1997}, explaining its trend with metallicity. In a supernova, Si yield is less sensitive to progenitor mass, so \xfe{Si} trends, in combination with \xfe{Mg}, can be used to probe the initial mass function (IMF) of stellar populations \citep{mcwilliam_chemistry_2013,hasselquist_apogee_2017}. In MWM DR19, Si is among the most precisely measured elements \citep{meszaros_sdss-v_2025}.

Compared to the ASPCAP raw \xfe{Si}-\feh plot in Figure \ref{fig:direct_feh_1}, our P4G LTE fits have a slight zero-point offset. We can see that the high-$\alpha$ track has a more gradual increase towards lower metallicities and does not appear to plateau. The P4G LTE low-$\alpha$ track also rises more steeply, in either direction from \feh=$-0.2$. For the P4G NLTE fits, the high-$\alpha$ track has a flatter trend, and the low-$\alpha$ track has a more steeply positive trend \feh$>-0.2$. 

We derive negative NLTE corrections of $\sim -0.04$ dex, as seen in Figure \ref{fig:nlte_corrs_0}. These are consistent with corrections derived in the NLTE Si analysis of \citet{zhang_nlte_2016}. NLTE does not reduce the temperature dependence of \xfe{Si} (Figure \ref{fig:occam_teff}).

NLTE has a similar magnitude effect on infrared Si abundances as on optical Si. \citet{amarsi_galah_2020} find that including NLTE mildly reduces the steepness of the \xfe{Si} trend, which is similar in our analysis, though their \xfe{Si} trends do not rise at supersolar metallicities, as ours do.

\subsection{Calcium (Ca)}
\label{subset:Ca}
Calcium is an $\alpha$-element, like Mg, and originates mostly from CCSNe, though has significant contribution from SNIa \citep{andrews_inflow_2017}. Like Si, Ca SN yield is less sensitive to progenitor mass, so \xfe{Ca} trends, in combination with \xfe{Mg}, can be used to probe the IMF of stellar populations \citep{mcwilliam_chemistry_2013,hasselquist_apogee_2017}.

We do not see particularly large NLTE differences for Ca. Compared to the raw ASPCAP values in Figure \ref{fig:direct_feh_1}, there is a slight zero-point offset of the low-$\alpha$ track. The P4G low-$\alpha$ track declines at super-solar metallicities, while the ASPCAP raw low-$\alpha$ track flattens out. The high-$\alpha$ track is extremely similar between the ASPCAP raw, P4G LTE, and P4G NLTE trends. As can be seen in Figure \ref{fig:nlte_corrs_1}, there appears to be relatively large star-to-star errors in our Ca fit, manifesting as a diffuse cloud around our corrections, though the average is consistent. Finally, Our NLTE correction lessens the trend of \xfe{Ca} on \teff, as can be seen in Figure \ref{fig:occam_teff}.

We compare our Ca NLTE correction to those derived by \citet{zhou_non-lte_2019} in Figure \ref{fig:nlte_corrs_1}. They generally agree with our corrections, though are slightly more positive. However, our corrections are similar to the zero-point offset applied in DR19 to bring Ca abundances in agreement with a solar neighborhood sample \citep{meszaros_sdss-v_2025}. Exclusion of NLTE could be a reason that the offset is required. We also compare our corrections to those generated by \citet{osorio_nlte_2020} for DR17 in Figure \ref{fig:osorio}. There is a clear systematic difference with temperature: our corrections agree at \teff=3750\,K and 4750\,K, though our corrections are more positive between those temperatures. 
Similar to Mg, \citet{osorio_nlte_2020} attributed differences between their work and \citet{zhou_non-lte_2019} to different collision data and also note that \code{MULTI} and \code{TLUSTY} \citep{hubeny_brief_2017} calculate background opacities in different manners. 

In our analysis and \citet{amarsi_galah_2020}, NLTE has a very little effect on Ca abundances. Their \xfe{Ca} trends show good agreement to our \xfe{Ca} trends.

\subsection{Titanium (Ti)}
\label{subsec:Ti}

Titanium is considered both an $\alpha$-element and an iron-peak element, with significant contributions from both CCSNe and SNIa \citep{andrews_inflow_2017}. In ASPCAP, its trend rises with metallicity, rather than the canonical $\alpha$-knee expected from an $\alpha$-element \citep{jonsson_apogee_2020, meszaros_sdss-v_2025}. \citet{hawkins_accurate_2016} point out that analysis of individual Ti I and II lines can result in discrepant trends, and they question if this effect is due to the exclusion of NLTE.

In our study, Ti is one of the elements most affected by NLTE. With the ASPCAP raw trend, we indeed see Ti increasing with increasing metallicity in Figure \ref{fig:direct_feh_1}. However, our P4G LTE Ti trend is quite flat for both the high- and low-$\alpha$ tracks. Our syntheses include Ti I and Ti II lines, without treating them as different species, but in ASPCAP, they report \code{ti\_h} as the abundance of neutral titanium. The inclusion of Ti II likely contributes to the difference between the ASPCAP raw and P4G LTE Ti trend. The P4G NLTE trend actually decreases with increasing metallicity.

Our NLTE trend appears to have the $\alpha$-knee, especially in the high-$\alpha$ track. The NLTE trend also exhibits a zero-point offset compared to the P4G LTE trend. 

NLTE does not lessen the dependence of \xfe{Ti} on \teff, which is one of the more profound \teff trends in our study. For Ti, our NN fits of our OCCAM sample have milder trends (not pictured), suggesting that the dependence comes from something specific to ASPCAP, such as blended lines in abundance windows.

\subsection{Manganese (Mn)}

Manganese is an iron-peak element, and is produced by both SNIa and CCSNe, alongside Fe. The positive trend of \xfe{Mn}-\feh \citep{andrews_inflow_2017, weinberg_chemical_2019} is thought to be a result of either metallicity-dependent production \citep{cescutti_chemical_2008} or delayed SNIa enrichment \citep{kobayashi_new_2020}. In SNIa, Mn is produced by nucleosynthetic pathways that are sensitive to the density of the progenitor white dwarf \citep{de_los_reyes_manganese_2020,kobayashi_new_2020}.

Mn exhibits strong differences when fit with NLTE. In Figure \ref{fig:direct_feh_1}, the P4G LTE \xfe{Mn}-\feh trend is less steep than the ASPCAP raw trend, and the P4G NLTE is even less steep. This results in a correction that varies along the red giant branch, as can be seen in Figure \ref{fig:nlte_corrs_1}. 

The corrections show a bimodality, which is related to the high- and low-$\alpha$ populations-- because they shift in different amounts and contribute to different regions of \feh-space, they cause the difference in the steepness of the \xfe{Mn}-\feh relationship. There is also a notably diffuse part of the P4G fits above \xfe{Mn}$>0.3$; this is driven by solar metallicity stars, and is likely due to poor fitting from molecular features.

Our NLTE correction lessens the dependence of \xfe{Mn} on \teff (Figure \ref{fig:occam_teff}), but does not eliminate it. Our P4G fits (not pictured) have similar trends to ASPCAP, so this could come from something common to both methods, like a lack of 3D modeling or the tuning of the linelist.

NLTE has a similar magnitude effect on infrared Mn abundances as on optical Si. \citet{amarsi_galah_2020} find that including NLTE reduces the steepness of the \xfe{Mn} trend, similar to our analysis. While \citet{bergemann_observational_2019} mostly discusses the effect of 3D NLTE in the derviation of Mn abundances, they also find that 1D LTE in H-band Mn lines underestimates 1D NLTE Mn abundance by between 0.05 and 0.2\, dex, which is consistent with our findings.

\subsection{Nickel (Ni)}

Nickel, like Mn, is an iron-peak element that is produced primarily by SNIa, with contribution from CCSNe \citep{kobayashi_role_2009,kobayashi_new_2020,andrews_inflow_2017}. It is also sensitive to the mass of the SNIa white dwarf progenitor \citep{kobayashi_new_2020}. However, unlike Mn, Ni has a relatively metallicity-independent trend, being produced \emph{in lockstep} with Fe \citep{hawkins_hunt_2023}. Ni is considered the most reliably determined iron-peak element in ASPCAP, aside from Fe itself \citep{meszaros_sdss-v_2025}.

Nickel exhibits modest differences when NLTE is included. Between the ASPCAP raw \xfe{Ni} and the P4G LTE \xfe{Ni} trends in Figure \ref{fig:direct_feh_1}, there is a slight zero-point offset, and again between the P4G LTE and NLTE. The diffuse part of the trend, around \feh=$-1$ can be associated with accreted populations of the Milky Way and changes slope between LTE and NLTE.

This manifests in the NLTE corrections in Figure \ref{fig:nlte_corrs_1}: the diffuse section, $\Delta$\xfe{Ni}$>0.1$ is mostly due to stars with \xh{Ni}$<-1.5$, corresponding to the chemically immature accreted population. Also, the NLTE correction lessens the trend of \xfe{Ni} with \teff (Figure \ref{fig:occam_teff}).

\section{Recommended Parameters and Corrections Catalogue}
\label{sec:vac}
Our analysis has formed the basis for a Value Added Catalogue (VAC) that is being released with SDSS DR20, \emph{apogee rgb nlte abundances}. The VAC consists of two files: \code{payne4GAIN\_modelSpectra\_v1.1.1.fits} and \code{payne4GAIN\_summary\_v1.1.1.fits}. The first file contains the spectra generated by our LTE and NLTE NNEs at the best fit parameters. The second file is a summary of the analysis we present in this paper. For each of the 360,451 stars, we include:
\begin{itemize}[nosep]
    \item raw (spectroscopic) data columns from ASPCAP: \teff, \logg, \vt, \feh, \xfe{X}. These begin with \code{aspcap\_raw\_}
    \item spectrum and result flags from ASPCAP: \teff, \logg, \vt, \feh, \xfe{X}. These are \code{aspcap\_...\_flags}
    \item Parameters from the LTE NN fit. These begin with \code{p4g\_lte\_}
    \item Parameters from the NLTE NN fit. These begin with \code{p4g\_nlte\_}
    \item ASPCAP raw abundances for the 8 elements we processed in NLTE, with the polynomial correction applied to them. These begin with \code{aspcap\_p4gcorr\_}
\end{itemize}
As part of the LTE and NLTE NN parameters, we include:
\begin{itemize}[nosep]
    \item Stellar labels: \teff, \logg, \vt, \feh, \xfe{C}, \xfe{N}, \xfe{O}, \xfe{Mg}, \xfe{Al}, \xfe{Si}, \xfe{Na}, \xfe{Mn}, \xfe{Ca}, \xfe{Ti}, \xfe{Ni}.
    \item The statistical errors from fitting. These are \code{p4g\_(n)lte\_...\_err}
    \item The chi-squared and degrees-of-freedom from fitting. These are \code{p4g\_(n)lte\_chi2} and \code{p4g\_(n)lte\_dof}
    \item The chi-squared of the model where \xfe{X} is set to its lowest value. This is to help with culling cases where fitting gives wildly varying results because the lines are small. These are \code{p4g\_(n)lte\_...\_uchi2}
    \item Manually assigned flags when the fit is close to the edge of the grid. These are \code{p4g\_(n)lte\_...\_gridflags}
\end{itemize}
The neural networks, as well as code examples for how to generate a spectrum at a given set of labels, can be found online\footnote{\url{https://github.com/pierrethx/Payne4GAIN}}.

The recommended ranges of our P4G-corrected abundances is provided in Table \ref{tab:limits}. The limits on \teff, \logg, and \xh{X} come from the ranges where the polynomial corrections agree with the trends of the star-by-star corrections, and correspond to the regions where the majority of our sample lies.
\begin{table}[h]
    \centering
    \caption{Recomended usage range of NLTE corrections}
    \begin{tabular}{c|c|c|c|c}
\hline
    Element & \teff & \logg & \xh{X} & Notes \\ \hline
    \xfe{Mg} & [3500,5250] & [1,3.5]& [-2,0.5] &\\
    \xfe{Si} & [3500,5250] & [1,3.5]& [-2,0.5] &\\
    \xfe{Al} & [3500,5250] &[1,3.5]& [-2,0.5] &\\
    \xfe{Ca} & [3500,5250] &[1,3.5]& [-1.5,0.5] &\\
    \xfe{Ti} &  [3500,5250] & [1,3.5]& [-1.5,1] &\\
    \xfe{Mn} & [3500,5250] & [1,3.5]& [-2,0.5] &\\
    \xfe{Ni} & [3500,5250] & [1,3.5]& [-2,0.5] &\\
    \xfe{Na} &  & & & Not recommended\\
\end{tabular}
    
    \label{tab:limits}
\end{table}

\section{Conclusion}
\label{sec:conclusion}
To summarize our NLTE results in the APOGEE H-band, we find strong NLTE effects for the IR lines of neutral ions of Aluminum, Titanium, and Manganese, consistent with the literature. These effects are of opposite signs: for Al I lines, the NLTE abundances are lower compared to LTE, whereas for Ti I and Mn I lines the abundances are higher by $\sim 0.1$ to $0.2$ dex compared to LTE, in agreement with previous work \citep[e.g.,][]{bergemann_ionization_2011, mallinson_titanium_2022,bergemann_observational_2019}. For Magnesium, Calcium, Silicon, and Nickel, we find mild NLTE corrections; our NLTE [Ni/Fe] results are very close to those presented by \citet{eitner_observational_2023}.  We also find that including NLTE in the chemical abundance analysis improves the scatter and tightens the chemical abundance trends [X/Fe], which is especially clear for [Ca/Fe] against [Fe/H] (Fig. \ref{fig:direct_feh_1}). Our method is unable to recover much useful information for Sodium.


In future data releases, we plan to improve the quality of the neural networks and the fitting, particularly the implementation of the line-spread function. These, in addition to the LTE-tuned linelist, are the primary impediments to recommending the use of our direct P4G fits. Because of these factors, for DR20 we derive polynomial-smoothed NLTE-LTE corrections and apply them to the ASPCAP raw values. 

The inclusion of NLTE Fe, which could have a profound effect on stellar parameters derived from H-band spectra, will require improved departure coefficient grids made specifically for infrared analysis. Future NLTE calculations should also keep H-band and other infrared lines in mind when determining which states are included in the model atoms. 

\section*{Acknowledgments}

P.N.T. and A.P.J. were supported by the National Science Foundation NSF AST-2206264 and NSF AST-2510795, and the University of Chicago's Research Computing Center for their support. A.P.J. acknowledges the Alfred P. Sloan Research Fellowship.
A.P.J. acknowledges the support of the NSF-Simons AI-Institute for the Sky (SkAI) via grants NSF AST-2421845 and Simons Foundation MPS-AI-00010513.

MB is supported from the European Research Council (ERC) under the European Union’s Horizon 2020 research and innovation programme (Grant agreement No. 949173).

Funding for the Sloan Digital Sky Survey V has been provided by the Alfred P. Sloan Foundation, the Heising-Simons Foundation, the National Science Foundation, and the Participating Institutions. SDSS acknowledges support and resources from the Center for High-Performance Computing at the University of Utah. SDSS telescopes are located at Apache Point Observatory, funded by the Astrophysical Research Consortium and operated by New Mexico State University, and at Las Campanas Observatory, operated by the Carnegie Institution for Science. The SDSS web site is \url{www.sdss.org}.

SDSS is managed by the Astrophysical Research Consortium for the Participating Institutions of the SDSS Collaboration, including the Carnegie Institution for Science, Chilean National Time Allocation Committee (CNTAC) ratified researchers, Caltech, the Gotham Participation Group, Harvard University, Heidelberg University, The Flatiron Institute, The Johns Hopkins University, L'Ecole polytechnique f\'{e}d\'{e}rale de Lausanne (EPFL), Leibniz-Institut f\"{u}r Astrophysik Potsdam (AIP), Max-Planck-Institut f\"{u}r Astronomie (MPIA Heidelberg), Max-Planck-Institut f\"{u}r Extraterrestrische Physik (MPE), Nanjing University, National Astronomical Observatories of China (NAOC), New Mexico State University, The Ohio State University, Pennsylvania State University, Smithsonian Astrophysical Observatory, Space Telescope Science Institute (STScI), the Stellar Astrophysics Participation Group, Universidad Nacional Aut\'{o}noma de M\'{e}xico, University of Arizona, University of Colorado Boulder, University of Illinois at Urbana-Champaign, University of Toronto, University of Utah, University of Virginia, Yale University, and Yunnan University.

This work has made use of the VALD database, operated at Uppsala University, the Institute of Astronomy RAS in Moscow, and the University of Vienna.

This work benefited from a workshop supported by the National Science Foundation under Grant No. OISE-1927130 (IReNA), the Kavli Institute for Cosmological Physics, and the University of Chicago Data Science Institute. 
This material is based upon work supported in part by the U.S. Department of Energy, Office of Science, Office of Nuclear Physics, under Award Number DE-SC0026204 and DE-SC0023128 (CeNAM).

J.G.F-T gratefully acknowledges the support provided by ANID Fondecyt Regular No. 1260371.

R.E. acknowledges support
from NSF grant AST-2206263, and NASA
ATP grant 80NSSC24K0899.

E.J.G. acknowledges support for this work provided by NASA through the NASA Hubble Fellowship Program grant No. HST-HF2-51576.001-A awarded by the Space Telescope Science Institute, which is operated by the Association of Universities for Research in Astronomy, Inc., for NASA, under the contract NAS 5-26555.



\bibliography{main}{}
\bibliographystyle{aasjournal}

\end{document}